\documentclass[english]{IEEEtran}
\usepackage[T1]{fontenc}
\usepackage[latin9]{inputenc}
\pdfoutput=1
\usepackage{geometry}
\usepackage{verbatim}
\usepackage{amsmath}
\usepackage{amsthm}
\usepackage{amssymb}
\usepackage{stackrel}
\usepackage{graphicx}

\makeatletter

\providecommand{\tabularnewline}{\\}

\theoremstyle{definition}
\newtheorem{assumption}{Assumption}
\theoremstyle{remark}
\newtheorem{rem}{\protect\remarkname}
\theoremstyle{definition}
\newtheorem{example}{\protect\examplename}
\theoremstyle{definition}
\newtheorem{defn}{\protect\definitionname}
\theoremstyle{plain}
\newtheorem{lem}{\protect\lemmaname}
\theoremstyle{plain}
\newtheorem{thm}{\protect\theoremname}

\usepackage{cite}
\usepackage{multirow}

\makeatother

\usepackage{babel}
\providecommand{\definitionname}{Definition}
\providecommand{\examplename}{Example}
\providecommand{\lemmaname}{Lemma}
\providecommand{\remarkname}{Remark}
\providecommand{\theoremname}{Theorem}

\begin{document}
\title{System Identification and Control Using Lyapunov-Based Deep Neural
Networks without Persistent Excitation: A Concurrent Learning Approach}
\author{Rebecca G. Hart, Omkar Sudhir Patil, Zachary I. Bell, and Warren E.
Dixon\thanks{This research is supported in part by the National Science Foundation
Graduate Research Fellowship under Grant No. DGE-2236414, AFRL grant
FA8651-24-1-0018, and AFOSR grant FA9550-19-1-0169. Any opinions,
findings, and conclusions or recommendations expressed in this material
are those of the author(s) and do not necessarily reflect the views
of the sponsoring agency.}\thanks{Rebecca G. Hart, Omkar Sudhir Patil, and Warren E. Dixon are with
the Department of Mechanical and Aerospace Engineering, University
of Florida, Gainesville, FL 32611 Emails: \{rebecca.hart, patilomkarsudhir,
wdixon\}@ufl.edu.}\thanks{Zachary I. Bell is with the Munitions Directorate, Air Force Research
Laboratory, Eglin AFB, FL 32542 Email: zachary.bell.10@us.af.mil}
\global\long\def\rr{\mathbb{R}}%
}
\maketitle
\begin{abstract}
Deep Neural Networks (DNNs) are increasingly used in control applications
due to their powerful function approximation capabilities. However,
many existing formulations focus primarily on tracking error convergence,
often neglecting the challenge of identifying the system dynamics
using the DNN. This paper presents the first result on simultaneous
trajectory tracking and online system identification using a DNN-based
controller, without requiring persistent excitation. Two new concurrent
learning adaptation laws are constructed for the weights of all the
layers of the DNN, achieving convergence of the DNN's parameter estimates
to a neighborhood of their ideal values, provided the DNN's Jacobian
satisfies a finite-time excitation condition. A Lyapunov-based stability
analysis is conducted to ensure convergence of the tracking error,
weight estimation errors, and observer errors to a neighborhood of
the origin. Simulations performed on a range of systems and trajectories,
with the same initial and operating conditions, demonstrated 40.5\%
to 73.6\% improvement in function approximation performance compared
to the baseline, while maintaining a similar tracking error and control
effort. Simulations evaluating function approximation capabilities
on data points outside of the trajectory resulted in 58.88\% and 74.75\%
improvement in function approximation compared to the baseline.
\end{abstract}

\section{Introduction}

Deep Neural Networks (DNNs) have become increasingly popular due to
their powerful function approximation capabilities \cite{Sun.Greene.ea2021,Joshi.Virdi.ea2020,Patil.Le.ea2022,Patil.Griffis.ea2023,Sweatland.Patil.ea2025,Basyal.Ting2024,Hart.Griffis.ea2024,Griffis.Patil.ea23_2,Shi.Shi.ea2019,Punjani.Abbeel2015,Bansal.Akametalu.ea2016,Li.Qian.ea2017,Karg2020}.
However, typical DNN-based estimators use offline optimization methods
to train the network weights. Such approaches require large data sets
and do not provide online persistent learning of unknown dynamics
which is relevant in applications which experience changes in the
underlying dynamics over time such as those involving smart materials,
fluid structure interaction, or human-machine interactions. Motivated
by these issues, \cite{Sun.Greene.ea2021} and \cite{Joshi.Virdi.ea2020}
introduced a DNN-based architecture which allows the outer layer weights
to adapt in real time while the inner layer weights are held static
with periodic updates using offline training techniques. Recent advancements
in results such as \cite{Patil.Le.ea2022,Hart.patil.ea2023,Le.Patil.ea.2022a,Griffis.Patil.ea23_2,Hart.Griffis.ea2024,Nino.Patil.ea2023,Lu.Wu.2024,Mei.Jian.2024}
(termed Lyapunov-based (Lb)-DNNs) have enabled online weight adaptation
for all the weights in a DNN, demonstrating improved tracking performance
over shallow NN approaches or traditional offline training methods.
However, these approaches have primarily focused on tracking objectives
and not parameter convergence or system identification. System identification
is an important objective because it allows the identified model to
provide insights that could be used in a variety of applications,
such as fault detection, or the ability to predict accurate state
estimates during intermittent loss of state feedback \cite{Pulido2024}.

For systems involving linear-in-parameter (LIP) uncertainty, results
such as \cite{Pan.Yu2018,Chowdhary.Muehlegg.ea2013} and \cite{Parikh.Kamalapurkar.ea2019,Kamalapurkar.Reish.ea2017,Roy.Bhasin.ea2017}
have developed methods to achieve parameter convergence without requiring
persistent excitation (PE), by incorporating data from previously
explored trajectories into the parameter update laws. The approach
in \cite{Basyal.Ting2024} used a LIP representation of a DNN by allowing
the output layer weights to adapt in real time and held the inner
layer weights static and applied the standard concurrent learning
(CL) method on the output layer weights. CL techniques, construct
a history stack of past information, resulting in improved system
identification as well as enabling off-trajectory exploration \cite{Kamalapurkar.Reish.ea2017,Parikh.Kamalapurkar.ea2019}.
This capability reduces over-reliance on instantaneous observations
and includes a wider range of past observations in the update process,
reducing the risk of converging to local minima. However, the nonlinear-in
parameter (NIP) structure of a DNN's inner layer presents challenges
in extending the CL update to all layers of the DNN making it an open
problem.

The system identification objective is difficult to achieve using
DNNs because they have nested nonlinearities and a NIP structure.
While classical adaptive control methods are known to achieve parameter
convergence for LIP uncertainties under the persistence of excitation
(PE) condition, only a few works consider NIP uncertainties \cite{Patil.Griffis.ea2023,Sweatland.Patil.ea2025,Pan.Yu2018,Basyal.Ting2024,Ortega.Aranovskiy.2021,Chowdhary.Muehlegg.ea2013}.
The result in \cite{Cao.Annaswamy.2003} considers convex/concave
nonlinear parameterizations to achieve global convergence of parameter
estimates, and the results in \cite{Wang.Ortega.ea2021} and \cite{Ortega.Bobtsov.2024}
consider strongly monotonic nonlinear parameterizations. However,
the NIP structure of DNNs is neither convex/concave nor strongly monotonic.

Motivated by the desire to achieve parameter convergence, results
such as \cite{Patil.Griffis.ea2023,Sweatland.Patil.ea2025,Ortega.Aranovskiy.2021,Pan.Yu2018,Basyal.Ting2024,Chowdhary.Muehlegg.ea2013}
use adaptive control techniques to yield convergence of parameter
estimates to a neighborhood of the actual parameters. In \cite{Patil.Griffis.ea2023}
and \cite{Sweatland.Patil.ea2025}, a prediction error formulation
was constructed using a dynamic state derivative estimator and a least
squares update law based on the DNN's Jacobian. This approach resulted
in parameter estimation error convergence to a neighborhood of the
origin, provided the Jacobian of the DNN satisfies the PE condition.
However, the PE condition on the Jacobian cannot be verified online
for nonlinear systems and requires continuous exploration of system
states and parameter trajectories; hence, posing a challenge in balancing
exploration and exploitation. Thus, online system identification
for DNNs without the PE condition remains an open problem.

This work presents the first approach for continuous all-layer adaptation
of a DNN informed by historical performance, using a CL-based update
law (hereafter referred to as Lb-CL-DNN). The use of the CL-based
update law relaxes the PE condition which is typical in other results
and yields convergence of the parameter estimates to a neighborhood
of the actual parameters under finite excitation (FE). Section \ref{sec:Continuous-Time-Nonlinear-Regres}
provides generalized identifiability conditions for nonlinear regression
equations (NREs) which are applicable to NIP models and demonstrates
that the FE condition is equivalent to ensuring that the unknown parameters
are identifiable. To overcome the challenges associated with the NIP
structure of the DNN, a first-order Taylor series approximation is
strategically applied at various recorded data-points. Then, two different
formulations are provided to construct a history stack using the DNN's
Jacobian and the Taylor series approximation applied on the data-points.
Section \ref{sec:Continuous-Time-Nonlinear-Regres} demonstrates the
update law on a continuous time regression problem and provides a
Lyapunov-based stability analysis to show the convergence of the weight
estimation errors. Sections \ref{sec:Unknown-System-Dynamics}-\ref{sec:Stability-AnalysisAdaptive}
apply the developed CL method to the adaptive control problem which
involves additional complexities due to the resulting regression involving
state-derivative terms which are typically noisy or unavailable. Therefore,
a dynamic state-derivative estimator is developed with a prescribable
settling time guarantee. By only collecting data after the prescribed
settling time, we prevent transient state estimation errors from corrupting
the history stack. Eliminating such transient errors is important
in this result because, unlike LIP regressions, the DNN's Jacobian
depends on the parameter estimates, and transient errors would result
in the history stack being filled with inaccurate estimates. To address
inaccuracies, two update laws were developed, each dynamically reconstructing
the history stack with newer parameter estimates and leveraging distinct
properties of DNNs to ensure accurate parameter estimation. A Lyapunov-based
stability analysis guarantees boundedness of the tracking error, weight
estimation errors, and observer errors. Simulations on multiple systems
and trajectories under the same operating conditions demonstrated
a 40.5\% to 73.6\% improvement in system identification compared to
the baseline, while maintaining a similar tracking error and control
effort. The improvement in function approximation capabilities was
also seen in simulations performed on off-trajectory data, with improvements
ranging from 58.88\% to 74.75\% .

\subsection*{Notation and Mathematical Background}

The space of essentially bounded Lebesgue measurable functions is
denoted by $\mathcal{L}_{\infty}$. Given $A\triangleq\left[a_{j,i}\right]\in\mathbb{R}^{n\times m},$
$\text{\text{vec}}(A)\triangleq\left[a_{1,1},\ldots,a_{n,1},\ldots,a_{1,m},\ldots,a_{n,m}\right]^{\top}$.
The Kronecker product is denoted by $\otimes$. Given any $A\in\rr^{n\times m}$,
$B\in\rr^{m\times p}$, and $C\in\rr^{p\times r}$, $\mathrm{vec}\left(ABC\right)=\left(C^{\top}\otimes A\right)\mathrm{vec}\left(B\right)$.
The right-to-left matrix product operator is represented by $\stackrel{\curvearrowleft}{\prod}$,
i.e., $\stackrel{\curvearrowleft}{\stackrel[p=1]{m}{\prod}}A_{p}=A_{m}\ldots A_{2}A_{1}$
and $\stackrel{\curvearrowleft}{\stackrel[p=a]{m}{\prod}}A_{p}=1$,
if $a>m$. The identity matrix of size $n\times n$ is denoted by
$I_{n}$. The zero vector of size $q\times1$ is given by $0_{q}$.
Given some functions $f$ and $g$ function composition is denoted
by $\circ$, where $\left(f\circ g\right)(x)\triangleq f(g(x)).$

\subsection{Deep Neural Network (DNN) Model\label{subsec:Deep-Neural-Network}}

DNNs are motivated given the prevalent evidence that indicates their
improved function approximation capabilities when compared to shallow
NNs \cite{Rolnick.Tegmark2018}. A feedforward DNN $\Phi\left(X,\theta\right)\in\mathbb{\mathbb{R}}^{L_{k+1}}$
can be modeled as \cite{Patil.Le.ea2022}
\begin{equation}
\Phi(X,\theta)\triangleq(v_{k}^{\top}\phi_{k}\circ...\circ v_{1}^{\top}\phi_{1})(v_{0}^{\top}X_{a}),\label{eq:GeneralDNN}
\end{equation}
where $\theta\triangleq\left[\text{\text{vec}}(v_{0})^{\top},...,\text{\text{vec}}(v_{k})^{\top}\right]^{\top}\in\mathbb{R}^{\rho}$,
where $\rho\triangleq\sum_{j=0}^{k}L_{j}L_{j+1},$ $j\in\left\{ 0,\dots,k\right\} $
and $k\in\mathbb{N}$ denotes the number of hidden layers within $\theta$,
$v_{j}\in\mathbb{R}^{L_{j}\times L_{j+1}}$ denotes the matrix of
weights and biases in the $j^{th}$ hidden layer, $L_{j}\in\mathbb{N}$
denotes the number of nodes within the $j^{th}$ hidden layer for
all $j\in\left\{ 0,...,k\right\} $, $k\in\mathbb{N}$ denotes the
number of hidden layers, and $L_{0}\triangleq m+1$, where $m$ is
the dimension of the $\mathbb{R}^{m}$ input vector to the DNN. Where
$X\in\Omega$ denotes the input to the DNN, $\Omega\subset\mathbb{R}^{m}$
denotes a compact set, and the augmented input $X_{a}\in\mathbb{R}^{m+1}$
is defined as $X_{a}\triangleq\begin{bmatrix}X^{\top} & 1\end{bmatrix}^{\top}$.
The vector of smooth activation functions at the $j^{th}$ layer is
denoted by $\phi_{j}\in\mathbb{R}^{L_{j}}$ and is defined as $\phi_{j}\triangleq\left[\varsigma_{j,1}\begin{array}{ccc}
\ldots & \varsigma_{j,L_{j}-1} & \mathrm{1}\end{array}\right]^{\top}$, where $\varsigma_{j,y}\in\mathbb{R}$ denotes the activation function
at the $y^{th}$ node of the $j^{th}$ layer for all $j\in\left\{ 1,...,k\right\} $.
To incorporate a bias term into the DNN model in (\ref{eq:GeneralDNN}),
the input $X$ and the activation functions $\phi_{j}$ are augmented
with a 1 for all $j\in\left\{ 1,...,k\right\} $. 

To facilitate the development of the online weight adaptation laws,
the DNN model in (\ref{eq:GeneralDNN}) can also be represented recursively
using the shorthand notation $\Phi_{j}$ as \cite{Patil.Le.ea2022}
\begin{equation}
\Phi_{j}\triangleq\begin{cases}
v_{j}^{\top}\phi_{j}(\Phi_{j-1}), & j\in\{1,...,k\},\\
v_{0}^{\top}X_{a} & j=0,
\end{cases}\label{eq:Phij_DNN}
\end{equation}
where $\Phi\left(X,\theta\right)=\Phi_{k}$.

The Jacobian of the feedforward DNN,\footnote{A fully-connected DNN model is considered for ease of exposition.
However, the CL update laws developed subsequently are agnostic to
the DNN architecture and can be applied to other architectures such
as \cite{Patil.Le.ea.2022,Griffis.Patil.ea23_2} based on their Jacobian
derivation.} denoted $\Phi^{\prime}(X,\theta)$, can be represented as $\Phi^{\prime}(X,\theta)\triangleq\left[\Phi_{0}^{\prime}(X,\theta),\ldots,\Phi_{j}^{\prime}(X,\theta)\right]\in\mathbb{R}^{n\times\rho}$,
where $\Phi_{j}^{\prime}\triangleq\frac{\partial\Phi_{j}(X,\theta)}{\partial\theta}\in\mathbb{R}^{n\times L_{j+1}}$,
for all $j\in\left\{ 0,\ldots,k\right\} $. Using (\ref{eq:Phij_DNN}),
the chain rule, and properties of the vectorization operator, the
terms $\Phi_{0}^{\prime}$ and $\Phi_{j}^{\prime}$ can be expressed
as \cite{Patil.Le.ea2022}{\footnotesize{}
\begin{align*}
\Phi_{0}^{\prime}(X,\hat{\theta}) & \triangleq\left(\stackrel{\curvearrowleft}{\prod_{l=1}^{k}}v_{l}^{\top}\phi_{l}^{\prime}\right)\left(I_{L_{1}}\otimes X_{a}^{\top}\right),\\
\Phi_{j}^{\prime} & (X,\theta)\triangleq\left(\stackrel{\curvearrowleft}{\prod_{l=j+1}^{k}}v_{l}^{\top}\phi_{l}^{\prime}\right)\left(I_{L_{j+1}}\otimes\phi_{j}^{\top}\right),
\end{align*}
}for all $j\in\left\{ 1,\ldots,k\right\} $, where the activation
function at the $j^{\text{th}}$ layer and its Jacobian are expressed
using the shorthand notations $\phi_{j}\triangleq\phi_{j}\left(\Phi_{j-1}\left(X,\theta\right)\right)$
and $\phi_{j}^{\prime}\triangleq\phi_{j}^{\prime}(\Phi_{j-1}(X,\theta))$,
respectively, and $\phi_{j}^{\prime}:\mathbb{R}^{L_{j}}\rightarrow\mathbb{R}^{L_{j}\times L_{j}}$
is defined as $\phi_{j}^{\prime}(y)\triangleq\frac{\partial}{\partial\varrho}\phi_{j}(\varrho)|_{\varrho=y}$,
for all $y\in\mathbb{R}^{L_{j}}$. To facilitate the subsequent development
and analysis, the following assumption is made on the activation functions.
\begin{assumption}
\label{assm:activation bounds} For each $j\in\left\{ 0,\ldots,k\right\} $,
the activation function $\phi_{j}$, its Jacobian $\phi_{j}^{\prime}$,
and Hessian $\phi_{j}^{\prime\prime}\left(y\right)\triangleq\frac{\partial^{2}}{\partial y^{2}}\phi_{j}\left(y\right)$
are bounded as
\begin{eqnarray}
\left\Vert \phi_{j}\left(y\right)\right\Vert  & \leq & \mathfrak{a}_{1}\left\Vert y\right\Vert +\mathfrak{a}_{0},\nonumber \\
\left\Vert \phi_{j}^{\prime}\left(y\right)\right\Vert  & \leq & \mathfrak{b}_{0},\nonumber \\
\left\Vert \phi_{j}^{\prime\prime}\left(y\right)\right\Vert  & \leq & \mathfrak{c}_{0},\label{eq:Activation Bounds}
\end{eqnarray}
where $\mathfrak{a}_{0},\mathfrak{a}_{1},\mathfrak{b}_{0},\mathfrak{c}_{0}\in\mathbb{R}_{\geq0}$
are known constants.
\end{assumption}
\begin{rem}
\label{rem:activation bounds} Most activation functions used in practice
satisfy Assumption \ref{assm:activation bounds}. Specifically, sigmoidal
activation functions (e.g., logistic function, hyperbolic tangent
etc.) have $\left\Vert \phi_{j}\left(y\right)\right\Vert $, $\left\Vert \phi_{j}^{\prime}\left(y\right)\right\Vert $,
and $\left\Vert \phi_{j}^{\prime\prime}\left(y\right)\right\Vert $
bounded uniformly by constants. Smooth approximations of rectified
linear unit (ReLUs) such as Swish grow linearly, and hence satisfy
the bound $\left\Vert \phi_{j}\left(y\right)\right\Vert \leq\mathfrak{a}_{1}\left\Vert y\right\Vert +\mathfrak{a}_{0}$
of Assumption \ref{assm:activation bounds}.
\end{rem}

\section{Continuous-Time Nonlinear Regression\label{sec:Continuous-Time-Nonlinear-Regres}}

Various applications spanning across the domains of system identification,
fault detection, financial time-series prediction, neuroscience, environmental
predictions, medical diagnostics etc. require online identification
of the underlying processes. Many such processes can be represented
as a continuous-time nonlinear regression. Therefore, consider the
NRE,
\begin{eqnarray}
y(t) & = & f(x(t))+\delta(t)\label{eq:mapping f}
\end{eqnarray}
for all $t\in\mathbb{R}_{\geq0}$, where $x(t)\in\Omega$ denotes
a known input to the regression and lies in the compact set $\Omega\subset\mathbb{R}^{m}$,
$y(t)\in\mathbb{R}^{n}$ denotes the output of the regression with
available measurements, $f:\Omega\to\mathbb{R}^{n}$ denotes an unknown
continuously differentiable function, and $\delta(t)\in\mathbb{R}^{n}$
denotes a bounded disturbance with known bound $\bar{\delta}\in\mathbb{R}_{>0}$
such that $\left\Vert \delta(t)\right\Vert \leq\bar{\delta}$ for
all $t\in\mathbb{R}_{\geq0}$. For notational brevity, time-dependencies
on all signals will be suppressed in the subsequent development. Because
the function $f$ is unknown and has no known structure, there are
challenges in learning the NRE. DNNs are a powerful tool for approximating
such unknown functions over compact sets, due to their universal function
approximation property \cite{Kidger.Lyons2020}. Therefore, a consider
a DNN $\Phi:\mathbb{R}^{m}\times\mathbb{R}^{\rho}\to\mathbb{R}^{n}$.
Then a DNN-based approximation of (\ref{eq:mapping f}) is given by
\begin{eqnarray}
\hat{f} & = & \Phi\left(x,\hat{\theta}\right),\label{eq:y hat}
\end{eqnarray}
where $\hat{\theta}\in\mathbb{R}^{\rho}$ denotes the DNN parameter
estimates that are designed based on the subsequent Lyapunov-based
stability analysis. The objective is to find the best estimates of
$\hat{\theta}$ such that $x\mapsto\Phi\left(x,\hat{\theta}\right)$
approximates $x\mapsto f(x)$ for any $x\in\Omega$. To quantify the
DNN-based nonlinear regression objective, let the loss function $\mathcal{L}:\mathbb{R}^{\rho}\to\mathbb{R}_{\geq0}$
be defined as
\begin{equation}
\mathcal{L}\left(\theta\right)\triangleq\int_{\Omega}\left(\left\Vert f\left(x\right)-\Phi\left(x,\theta\right)\right\Vert ^{2}+\sigma\left\Vert \theta\right\Vert ^{2}\right)d\mu\left(x\right),\label{eq:Loss Function}
\end{equation}
where $\mu$ denotes the Lebesgue measure, $\sigma\in\mathbb{R}_{>0}$
denotes a regularizing constant, and the term $\sigma\left\Vert \theta\right\Vert ^{2}$
represents $L_{2}$ regularization (also popularly known as ridge
regression in the machine learning community) \cite[Sec. 7.1.1]{Goodfellow.Bengio.ea2016}.
Note that this loss function is defined independently of the time-dependent
signal $t\mapsto x(t)$ because the objective is to ensure the approximation
holds for any $x\in\Omega$ and not just the set of points traversed
by $t\mapsto x(t)$. Additionally, a user-selected compact convex
parameter search space $\Theta$ satisfying $0_{\rho}\in\Theta$ and
having a smooth boundary is considered. Additionally, $\bar{\theta}\triangleq\underset{\theta\in\Theta}{\max}\left\Vert \theta\right\Vert \in\mathbb{R}_{>0}$
is a bound on the user-defined search space. The objective is to identify
the vector of ideal DNN parameters $\theta^{*}\in\Theta$ defined
as
\begin{eqnarray}
\theta^{*} & \triangleq & \underset{\theta\in\Theta}{\arg\min}\ \mathcal{L}\left(\theta\right).\label{eq:theta star}
\end{eqnarray}

Although using a bounded search space can restrict the optimality
of the identified parameters to be local instead of global, it allows
the subsequent development to be analyzed from a convex optimization
perspective, which otherwise would be non-convex due to the nested
NIP structure of DNNs. Specifically, due to the strict convexity of
the regularizing term $\sigma\left\Vert \theta\right\Vert ^{2}$ in
(\ref{eq:Loss Function}), there exists $\sigma\in\mathbb{R}_{>0}$
which ensures $\mathcal{L}\left(\theta\right)$ is convex for all
$\theta\in\Theta$. Additionally, the regularizing term has other
advantages such as mitigation of overfitting \cite[Sec. 7.1.1]{Goodfellow.Bengio.ea2016}.
However, selecting very high values of $\sigma$ can obscure the contribution
of the $\left\Vert f\left(x\right)-\Phi\left(x,\theta\right)\right\Vert ^{2}$
term to the loss function while also causing underfitting \cite[Sec. 7.1.1]{Goodfellow.Bengio.ea2016};
therefore, there is a tradeoff between selecting low vs. high values
of $\sigma$. Furthermore, note that that local minima for the $\left\Vert f\left(x\right)-\Phi\left(x,\theta\right)\right\Vert ^{2}$
term are ubiquitous for various applications of DNNs. The important
question is whether there are local minima of higher cost than the
global minima. As noted in \cite[Sec. 8.2.2.]{Goodfellow.Bengio.ea2016},
this question remains open for general DNN architectures. However,
for some DNN architectures, it has been established that every local
minima is a global minima. For example, \cite{Kawaguchi.Bengio2019}
established this property for deep residual neural networks (ResNets)
with arbitrary nonlinear activation functions. Results such as \cite{Kawaguchi2016}
and \cite{Lu.Kawaguchi2017} have concluded the same for DNNs with
linear activation functions, and \cite{Du.Lee.ea2018} establishes
the same for single-hidden-layer deep convolutional neural networks
with rectified linear unit (ReLU) activation functions. Furthermore,
from a practical standpoint, it is acceptable to find a point in the
parameter space that has low but not necessarily minimal cost \cite[Sec. 8.2.2.]{Goodfellow.Bengio.ea2016}.
Therefore, we restrict our attention to the problem of finding a local
minima. However, even this problem is challenging because a DNN parameterization
can be represented using multiple equivalent parameterizations. 
\begin{example}
Consider a scalar input-output NN with one hidden linear and one hidden
neuron, $\Phi\left(x,\theta\right)=w_{1}w_{0}x$, where $w_{1}$ and
$w_{0}$ are scalar NN weights. Then, for any $\alpha\in\mathbb{R}_{>0}$,
defining the new weights $v_{1}=\alpha w_{1}$ and $v_{0}=\frac{w_{0}}{\alpha}$
yields an equivalent parameterization $\Phi\left(x,\theta\right)=v_{1}v_{0}x$
with different weights.
\end{example}
This equivalence leads to model identifiability issues, where the
presence of uncountably infinite indistinguishable local minima can
make the problem of identifying $\theta^{*}$ ill-posed \cite[Sec. 8.2.2.]{Goodfellow.Bengio.ea2016}.
To address this issue, the following subsection provides identifiability
conditions for DNNs. 

\subsection{Identifiability Conditions}

To define the conditions under which the problem of identifying $\theta^{*}$
is well-posed, we provide a definition of least squares identifiability,
which is a modified version of \cite[Definition 3]{Grewal.Glover1976}
tailored to the regression problem in (\ref{eq:mapping f}). 
\begin{defn}
\label{def:Identifiability}The parameter vector $\theta^{*}$ is
identifiable over the set $\Theta$ if and only if it is a unique
minimizer of $\mathcal{L}$ over $\Theta$. If $\theta^{*}$ is only
an isolated local minimizer (i.e., if there exists an arbitrarily
small neighborhood of $\theta^{*}$ where it is a unique minimizer),
then $\theta^{*}$ is termed locally identifiable. If $\theta^{*}$
is identifiable over the entire Euclidean space $\mathbb{R}^{\rho}$,
then it is globally identifiable.
\end{defn}
For general deep learning tasks, achieving global identifiability
can be intractable due to the aforementioned reasons for why we consider
a bounded search space; therefore, we examine conditions such that
$\theta^{*}$ is locally identifiable or identifiable over $\Theta$.
To obtain the conditions under which $\theta^{*}$ is identifiable
over $\Theta$, strict convexity is imposed on $\mathcal{L}$. Because
$\Theta$ is a convex set, imposing strict convexity on $\mathcal{L}$
ensures unique solutions to (\ref{eq:theta star}). A sufficient condition
for ensuring strict convexity is ensuring the Hessian $\frac{\partial^{2}\mathcal{L}}{\partial\theta^{2}}$
is positive definite, i.e., $\frac{\partial^{2}\mathcal{L}}{\partial\theta^{2}}\succ0$
for all $\theta\in\Theta$ \cite[Sec. 3.1.4]{Boyd2004}. Imposing
positive-definiteness on $\frac{\partial^{2}\mathcal{L}}{\partial\theta^{2}}$
yields the condition
\begin{eqnarray}
\int_{\Omega}\left(\Phi^{\prime\top}\left(x,\theta\right)\Phi^{\prime}\left(x,\theta\right)+\sigma I_{\rho}\right)d\mu\left(x\right)\nonumber \\
-\int_{\Omega}\frac{\partial^{2}\Phi\left(x,\theta\right)}{\partial\theta^{2}}\left(f\left(x\right)-\Phi\left(x,\theta\right)\right)d\mu\left(x\right) & \succ & 0,\label{eq:Identifiability under strict convexity}
\end{eqnarray}
for all $\theta\in\Theta$. Although the inequality in (\ref{eq:Identifiability under strict convexity})
offers an identifiability condition, this condition is challenging
to verify because it involves a partial integro-differential inequality
consisting of the unknown function approximation error $f\left(x\right)-\Phi\left(x,\theta\right)$
and the second-derivative of the DNN $\frac{\partial^{2}\Phi\left(x,\theta\right)}{\partial\theta^{2}}$
which is often computationally intractable. However, stricter conditions
can be imposed to obtain tractable sufficient conditions for identifiability.
To derive these stricter conditions, notice the term $\frac{\partial^{2}\Phi\left(x,\theta\right)}{\partial\theta^{2}}\left(f\left(x\right)-\Phi\left(x,\theta\right)\right)$
is bounded for all $x\in\Omega$ and $\theta\in\Theta$; this fact
can be established by noticing the terms $\frac{\partial^{2}\Phi\left(x,\theta\right)}{\partial\theta^{2}}$
and $f\left(x\right)-\Phi\left(x,\theta\right)$ are bounded for all
$x\in\Omega$ and $\theta\in\Theta$ due to $\Phi$ being $\mathcal{C}^{2}$
and $f$ being continuous. Therefore, let $\bar{\sigma}\in\mathbb{R}_{>0}$
denote the bound $\bar{\sigma}\triangleq\underset{x\in\Omega,\theta\in\Theta}{\sup}\left\Vert \frac{\partial^{2}\Phi\left(x,\theta\right)}{\partial\theta^{2}}\left(f\left(x\right)-\Phi\left(x,\theta\right)\right)\right\Vert $.
Then, a stricter condition for identifiability is obtained as 
\begin{equation}
\int_{\Omega}\left(\Phi^{\prime\top}\left(x,\theta\right)\Phi^{\prime}\left(x,\theta\right)+\sigma I_{\rho}-\bar{\sigma}I_{\rho}\right)d\mu\left(x\right)\succ0.\label{eq:Relatively stricter identifiability}
\end{equation}

\begin{rem}
\label{rem:SigmaMod}Selecting a large regularizing constant $\sigma$
can trivially ensure the condition in (\ref{eq:Relatively stricter identifiability}).
However, doing so is undesirable as it obscures the DNN's contribution
to the loss function in $\eqref{eq:Loss Function}$. Instead, it is
desirable that the term $\int_{\Omega}\Phi^{\prime\top}\left(x,\theta\right)\Phi^{\prime}\left(x,\theta\right)d\mu\left(x\right)$
contributes to achieving the identifiability condition in (\ref{eq:Relatively stricter identifiability}),
which yields the condition $\int_{\Omega}\Phi^{\prime\top}\left(x,\theta\right)\Phi^{\prime}\left(x,\theta\right)d\mu\left(x\right)\succ\bar{\sigma}I_{\rho}\mu\left(\Omega\right)$.%
\begin{comment}
Wouldn't this be $(\overline{\sigma}-\sigma)I_{\rho}\mu(\Omega)$?
\end{comment}
{} Note that this condition, although still computationally intensive,
is relatively easier to verify using a discrete approximation of the
integral as compared to (\ref{eq:Relatively stricter identifiability}).
Alternatively, a more conservative approach can be taken where $\sigma$
is selected to be equal to $\bar{\sigma}$ while requiring only positive-definiteness
to be verified for $\int_{\Omega}\Phi^{\prime\top}\left(x,\theta\right)\Phi^{\prime}\left(x,\theta\right)d\mu\left(x\right)$. 

The following lemma provides equivalent conditions for verifying if
$\int_{\Omega}\Phi^{\prime\top}\left(x,\theta\right)\Phi^{\prime}\left(x,\theta\right)d\mu\left(x\right)$
is positive definite to check for identifiability.
\end{rem}
\begin{lem}
\label{lem:identifiability}Let $\sigma=\bar{\sigma}$. Then $\theta^{*}$
is identifiable over $\Theta$ if any of the following equivalent
conditions are satisfied,

i)$\int_{\Omega}\Phi^{\prime\top}\left(x,\theta\right)\Phi^{\prime}\left(x,\theta\right)d\mu\left(x\right)\succ0$
for all $\theta\in\Theta$,%
\begin{comment}
Agreed
\end{comment}

ii) there exists $x_{1},\ldots,x_{\rho}\in\Omega$ such that $\sum_{i=1}^{\rho}\Phi^{\prime\top}\left(x_{i},\theta\right)\Phi^{\prime}\left(x_{i},\theta\right)\succ0$
for all $\theta\in\Theta$,%
\begin{comment}
Agreed--is this putting a condition on the history stack such that
it has to be at least as large as the number of parameters?
\end{comment}

iii) there exists $x_{1},\ldots,x_{\rho}\in\Omega$ such that $\textrm{rank}\left(\left[\Phi^{\prime\top}\left(x_{1},\theta\right),\ldots,\Phi^{\prime\top}\left(x_{p},\theta\right)\right]\right)=\rho$
for all $\theta\in\Theta$.%
\begin{comment}
Agreed
\end{comment}
\end{lem}
\begin{IEEEproof}
Condition i) follows from (\ref{eq:Relatively stricter identifiability})
as stated before. By the definition of positive-definiteness of a
matrix, condition i) is equivalent to stating $\int_{\Omega}v^{\top}\Phi^{\prime\top}\left(x,\theta\right)\Phi^{\prime}\left(x,\theta\right)vd\mu\left(x\right)=\int_{\Omega}\left\Vert \Phi^{\prime}\left(x,\theta\right)v\right\Vert ^{2}d\mu\left(x\right)>0$
for all $v\neq0_{\rho}$.%
\begin{comment}
Agreed--I feel like this proves i)
\end{comment}
{} Furthermore, consider an arbitrary partition of $\Omega$ formed
by sets $\Omega_{1},\ldots,\Omega_{\rho}\subset\Omega$. Then the
integral $\int_{\Omega}\left\Vert \Phi^{\prime}\left(x,\theta\right)v\right\Vert ^{2}d\mu\left(x\right)$
can be upper-bounded as $\int_{\Omega}\left\Vert \Phi^{\prime}\left(x,\theta\right)v\right\Vert ^{2}d\mu\left(x\right)\leq\sum_{i=1}^{\rho}\left(\underset{x\in\Omega_{i}}{\sup}\left\Vert \Phi^{\prime}\left(x,\theta\right)v\right\Vert ^{2}\right)\mu\left(\Omega_{i}\right)$.%
\begin{comment}
Yeah this seems fair
\end{comment}
{} Since $\underset{x\in\Omega_{i}}{\sup}\left\Vert \Phi^{\prime}\left(x,\theta\right)v\right\Vert ^{2}$
and $\mu\left(\Omega_{i}\right)$ are non-negative,%
\begin{comment}
Agreed
\end{comment}
{} the Cauchy-Schwarz inequality can be applied which yields $\sum_{i=1}^{\rho}\left(\underset{x\in\Omega_{i}}{\sup}\left\Vert \Phi^{\prime}\left(x,\theta\right)v\right\Vert ^{2}\right)\mu\left(\Omega_{i}\right)\leq\left(\sum_{i=1}^{\rho}\underset{x\in\Omega_{i}}{\sup}\left\Vert \Phi^{\prime}\left(x,\theta\right)v\right\Vert ^{2}\right)\left(\sum_{i=1}^{\rho}\mu\left(\Omega_{i}\right)\right)$.%
\begin{comment}
The Cauchy-Schwarz inequality states that $\left(\sum_{i=1}^{n}u_{i}v_{i}\right)^{2}\leq\left(\sum_{i=1}^{n}u_{i}^{2}\right)\left(\sum_{i=1}^{n}v_{i}^{2}\right)$
\end{comment}
\begin{comment}
Agreed
\end{comment}
{} Hence, condition i) implies $\left(\sum_{i=1}^{\rho}\underset{x\in\Omega_{i}}{\sup}\left\Vert \Phi^{\prime}\left(x,\theta\right)v\right\Vert ^{2}\right)\left(\sum_{i=1}^{\rho}\mu\left(\Omega_{i}\right)\right)>0$
for all $v\neq0_{\rho}$%
\begin{comment}
Previously Stated: 
\begin{align*}
0 & <\int_{\Omega}v^{\top}\Phi^{\prime\top}\left(x,\theta\right)\Phi^{\prime}\left(x,\theta\right)vd\mu\left(x\right)=\int_{\Omega}\left\Vert \Phi^{\prime}\left(x,\theta\right)v\right\Vert ^{2}d\mu\left(x\right)\\
 & \leq\sum_{i=1}^{p}\left(\underset{x\in\Omega_{i}}{\sup}\left\Vert \Phi^{\prime}\left(x,\theta\right)v\right\Vert ^{2}\right)\mu\left(\Omega_{i}\right)\\
 & \leq\left(\sum_{i=1}^{p}\underset{x\in\Omega_{i}}{\sup}\left\Vert \Phi^{\prime}\left(x,\theta\right)v\right\Vert ^{2}\right)\left(\sum_{i=1}^{p}\mu\left(\Omega_{i}\right)\right)
\end{align*}
so then yes $0<\left(\sum_{i=1}^{p}\underset{x\in\Omega_{i}}{\sup}\left\Vert \Phi^{\prime}\left(x,\theta\right)v\right\Vert ^{2}\right)\left(\sum_{i=1}^{p}\mu\left(\Omega_{i}\right)\right)$
\end{comment}
. Due to the property of partitions $\sum_{i=1}^{\rho}\mu\left(\Omega_{i}\right)=\mu\left(\Omega\right)>0$%
\begin{comment}
Agreed
\end{comment}
, the inequality $\sum_{i=1}^{\rho}\underset{x\in\Omega_{i}}{\sup}\left\Vert \Phi^{\prime}\left(x,\theta\right)v\right\Vert ^{2}>0$
is obtained%
\begin{comment}
Agreed- we did already say that $\underset{x\in\Omega_{i}}{\sup}\left\Vert \Phi^{\prime}\left(x,\theta\right)v\right\Vert ^{2}$
was non negative so this may be a little redundant since the sum of
non-negative elements would also be non-negative
\end{comment}
. Therefore, selecting $x_{i}=\underset{x\in\Omega_{i}}{\arg\sup}\left\Vert \Phi^{\prime}\left(x,\theta\right)v\right\Vert ^{2}$
yields $\sum_{i=1}^{\rho}\underset{x\in\Omega_{i}}{\sup}\left\Vert \Phi^{\prime}\left(x,\theta\right)v\right\Vert ^{2}=\sum_{i=1}^{\rho}v^{\top}\Phi^{\prime\top}\left(x_{i},\theta\right)\Phi^{\prime}\left(x_{i},\theta\right)v>0$
for all $v\neq0_{\rho}$, thus implying $\sum_{i=1}^{\rho}\Phi^{\prime\top}\left(x_{i},\theta\right)\Phi^{\prime}\left(x_{i},\theta\right)\succ0$
\begin{comment}
Agreed -- note to self that people will probably say that it will
be hard to find the $x_{i}'s$ that do this
\end{comment}
{} for all $\theta\in\Theta$. Hence, condition i) implies condition
ii). 

To establish that conditions i) and ii) are equivalent, we now establish
that condition i) is also necessary to satisfy condition ii). We establish
this by the method of contradiction. Thus, assume for the sake of
contraction that condition i) is violated and condition ii) is satisfied.
Due to the violation of condition i), $\int_{\Omega}\Phi^{\prime\top}\left(x,\theta\right)\Phi^{\prime}\left(x,\theta\right)d\mu\left(x\right)$
is not positive-definite but only positive semi-definite. In this
case, there exists $v\neq0_{\rho}$ such that $\int_{\Omega}\left\Vert \Phi^{\prime}\left(x,\theta\right)v\right\Vert ^{2}d\mu\left(x\right)=0$.
Furthermore, notice that $\int_{\Omega}\left\Vert \Phi^{\prime}\left(x,\theta\right)v\right\Vert ^{2}d\mu\left(x\right)$
can be lower-bounded as $0\leq\sum_{i=1}^{\rho}\left(\underset{x\in\Omega_{i}}{\inf}\left\Vert \Phi^{\prime}\left(x,\theta\right)v\right\Vert ^{2}\right)\mu\left(\Omega_{i}\right)\leq\int_{\Omega}\left\Vert \Phi^{\prime}\left(x,\theta\right)v\right\Vert ^{2}d\mu\left(x\right)$.
Therefore, if $\int_{\Omega}\left\Vert \Phi^{\prime}\left(x,\theta\right)v\right\Vert ^{2}d\mu\left(x\right)=0$,
then $\sum_{i=1}^{\rho}\left(\underset{x\in\Omega_{i}}{\inf}\left\Vert \Phi^{\prime}\left(x,\theta\right)v\right\Vert ^{2}\right)\mu\left(\Omega_{i}\right)=0$,
which is only possible when $\underset{x\in\Omega_{i}}{\inf}\left\Vert \Phi^{\prime}\left(x,\theta\right)v\right\Vert ^{2}=0$
for all $i\in\left\{ 1,\ldots,\rho\right\} $. Since the partitions
are arbitrary, this condition must apply for all possible partitions
of $\Omega$, i.e., there does not exist a method to partition $\Omega_{i}$
such that $\underset{x\in\Omega_{i}}{\inf}\left\Vert \Phi^{\prime}\left(x,\theta\right)v\right\Vert ^{2}>0$
for any $i\in\left\{ 1,\ldots,\rho\right\} $. Since condition ii)
is assumed to be satisfied, there exists $x_{1},\ldots,x_{\rho}\in\Omega$
such that $\sum_{i=1}^{\rho}\left\Vert \Phi^{\prime}\left(x_{i},\theta\right)v\right\Vert ^{2}>0$,
implying $\left\Vert \Phi^{\prime}\left(x_{i},\theta\right)v\right\Vert ^{2}>0$
for at least one $i\in\left\{ 1,\ldots,\rho\right\} $. For such an
$i$, due to the continuity of $\left\Vert \Phi^{\prime}\left(x_{i},\theta\right)v\right\Vert ^{2}$,
there exists a neighborhood of $x_{i}$ of radius $\epsilon\in\mathbb{R}_{>0}$
given by $\mathcal{N}_{i}\triangleq\left\{ x\in\Omega:\left\Vert x-x_{i}\right\Vert \leq\epsilon\right\} $
such that $\left\Vert \Phi^{\prime}\left(x,\theta\right)v\right\Vert ^{2}>0$
for all $x\in\mathcal{N}_{i}$. Additionally, note that $\mu\left(\mathcal{N}_{i}\right)>0$.
Then partitioning $\Omega$ such that $\Omega_{i}=\mathcal{N}_{i}$
yields $\underset{x\in\Omega_{i}}{\inf}\left\Vert \Phi^{\prime}\left(x,\theta\right)v\right\Vert ^{2}>0$,
thus leading to a contradiction with the implications of condition
i). Therefore, one cannot violate condition i) and satisfy condition
ii) simultaneously. Hence, condition i) is necessary and sufficient
for condition ii).%
\begin{comment}
Agreed
\end{comment}

Now we establish the equivalence between conditions ii) and iii).
Notice that condition ii) holds if and only if $\textrm{rank}\left(\sum_{i=1}^{\rho}\Phi^{\prime\top}\left(x_{i},\theta\right)\Phi^{\prime}\left(x_{i},\theta\right)\right)=\rho$,
which is equivalent to the statement $\ker\left(\sum_{i=1}^{\rho}\Phi^{\prime\top}\left(x_{i},\theta\right)\Phi^{\prime}\left(x_{i},\theta\right)\right)=\{0_{\rho}\}$
due to the rank-nullity theorem \cite[Thm. 2.3]{Friedberg.Insel.ea2003}.
Additionally, due to the property that $\ker\left(A+B\right)\supseteq\ker\left(A\right)\cap\ker\left(B\right)$
for any given matrices $A$ and $B$, the condition $\ker\left(\sum_{i=1}^{\rho}\Phi^{\prime\top}\left(x_{i},\theta\right)\Phi^{\prime}\left(x_{i},\theta\right)\right)=\{0_{\rho}\}$
holds if and only if $\bigcap_{i=1}^{\rho}\ker\left(\Phi^{\prime\top}\left(x_{i},\theta\right)\Phi^{\prime}\left(x_{i},\theta\right)\right)=\{0_{\rho}\}$.
Furthermore, note that $\ker\left(\Phi^{\prime\top}\left(x_{i},\theta\right)\Phi^{\prime}\left(x_{i},\theta\right)\right)=\ker\left(\Phi^{\prime}\left(x_{i},\theta\right)\right)$
which is established as follows. For any vector $v\in\mathbb{R}^{\rho}$,
if $\Phi^{\prime}\left(x_{i},\theta\right)v=0_{n}$, then multiplying
by $\Phi^{\prime\top}\left(x_{i},\theta\right)$ on both sides yields
$\Phi^{\prime\top}\left(x_{i},\theta\right)\Phi^{\prime}\left(x_{i},\theta\right)v=0_{\rho}$.
Additionally, if $\Phi^{\prime\top}\left(x_{i},\theta\right)\Phi^{\prime}\left(x_{i},\theta\right)v=0_{\rho}$,
then multiplying by $v^{\top}$ on both sides yields $v^{\top}\Phi^{\prime\top}\left(x_{i},\theta\right)\Phi^{\prime}\left(x_{i},\theta\right)v=0$,
equivalent to $\left\Vert \Phi^{\prime}\left(x_{i},\theta\right)v\right\Vert =0$
which holds if and only if $\Phi^{\prime}\left(x_{i},\theta\right)v=0_{n}$
\begin{comment}
Conflicting dimension with the next statement
\end{comment}
. Therefore, $\Phi^{\prime\top}\left(x_{i},\theta\right)\Phi^{\prime}\left(x_{i},\theta\right)v=0_{\rho}$
if and only if $\Phi^{\prime}\left(x_{i},\theta\right)v=0_{\rho}$%
\begin{comment}
conflicts with `` $\left\Vert \Phi^{\prime}\left(x_{i},\theta\right)v\right\Vert =0$
which holds if and only if $\Phi^{\prime}\left(x_{i},\theta\right)v=0_{n}$''
\end{comment}
, thus implying $\ker\left(\Phi^{\prime\top}\left(x_{i},\theta\right)\Phi^{\prime}\left(x_{i},\theta\right)\right)=\ker\left(\Phi^{\prime}\left(x_{i},\theta\right)\right)$.
Therefore, condition ii) is equivalent to the condition $\bigcap_{i=1}^{\rho}\ker\left(\Phi^{\prime}\left(x_{i},\theta\right)\right)=\{0_{\rho}\}$,
which holds if and only if there exists no vector $v\neq0_{\rho}$
such that $\Phi^{\prime}\left(x_{i},\theta\right)v=0_{n}$ %
\begin{comment}
Looks like the conflict was intentional
\end{comment}
for all $i\in\left\{ 1,\ldots,\rho\right\} $. This condition can
hold if and only if the matrix $\left[\Phi^{\prime\top}\left(x_{1},\theta\right),\ldots,\Phi^{\prime\top}\left(x_{\rho},\theta\right)\right]^{\top}$
has $\rho$ linearly independent columns, which is equivalent to the
statement in condition iii) that $\textrm{rank}\left(\left[\Phi^{\prime\top}\left(x_{1},\theta\right),\ldots,\Phi^{\prime\top}\left(x_{\rho},\theta\right)\right]\right)=\rho$.
Therefore, conditions ii) and iii) are equivalent, thus establishing
the equivalence between conditions i)-iii).
\end{IEEEproof}
\begin{rem}
\label{rem:local identifiability}To determine local identifiability
(i.e., identifiability in an arbitrarily small neighborhood of $\theta^{*}$
as defined in Definition \ref{def:Identifiability}) instead of identifiability
over $\Theta$, the conditions i)-iii) stated in Lemma \ref{lem:identifiability}
can be relaxed to hold only at $\theta^{*}$ instead of holding for
all $\theta\in\Theta$. Furthermore, $\bar{\sigma}$ can be redefined
as $\bar{\sigma}=\underset{x\in\Omega}{\sup}\left\Vert \frac{\partial^{2}\Phi\left(x,\theta^{*}\right)}{\partial\theta^{2}}\left(f\left(x\right)-\Phi\left(x,\theta^{*}\right)\right)\right\Vert $%
\begin{comment}
Note: not able to be calculated because $\theta^{*}$ which is unknown--might
not be an issue but just noting, unless you can calculate it a priori
with the loss fcn
\end{comment}
{} to check for local identifiability. This redefinition of $\bar{\sigma}$
is expected to yield a significantly smaller value than its previous
definition because it does not involve the supremum to be computed
over the entire $\Theta$ but only at $\theta^{*}$.
\end{rem}
\begin{rem}
\label{rem:Universal Function Approximation}Notice that the universal
function approximation theorem for DNNs was not invoked in the definition
of $\theta^{*}$ or in the derivation of the identifiability conditions.
The universal function approximation theorem \cite[Theorem 3.1]{Kidger.Lyons2020}
states that the function space of DNNs is dense in the space of continuous
functions $\mathcal{C}\left(\Omega\right)$. As a result, for any
prescribed $\varepsilon>0$, there exists a DNN $\Phi$ and a corresponding
parameter $\theta$ such that $\underset{x\in\Omega}{\sup}\left\Vert f\left(x\right)-\Phi\left(x,\theta\right)\right\Vert <\varepsilon$,
and therefore $\int_{\Omega}\left\Vert f\left(x\right)-\Phi\left(x,\theta\right)\right\Vert ^{2}d\mu\left(x\right)<\varepsilon^{2}\mu\left(\Omega\right)$.
However, it is not known how to obtain a bound $\bar{\theta}$ on
such parameter $\theta$ for an arbitrary $\varepsilon$, which causes
difficulties in constructing the bounded search space $\Theta$. Therefore,
we allow the user-defined search space $\Theta$ to be arbitrarily
selected in the above analysis, at the loss of guarantees on the approximation
accuracy. Although the constant $\varepsilon$ that bounds $\underset{x\in\Omega}{\sup}\left\Vert f\left(x\right)-\Phi\left(x,\theta{}^{*}\right)\right\Vert $
might no longer be arbitrary in this case, it would still be finite\textbf{
}due to the continuity of $f$ and $\Phi$, where minimizing (\ref{eq:Loss Function})
would yield the best regularized approximation of $f$. Therefore,
the unknown nonlinear function can be modeled as 
\begin{equation}
f(x)=\Phi(x,\theta^{*})+\varepsilon(x),\label{eq:UFA-1}
\end{equation}
 where $\varepsilon(x)\triangleq f(x)-\Phi(x,\theta^{*})$ is bounded
by $\overline{\varepsilon}$ as $\Vert\varepsilon(x)\Vert\leq\overline{\varepsilon}$
for all $x\in\Omega$.
\end{rem}
Recent work \cite{Wang.Ortega.ea2021} provided identifiability conditions
for linear regression equations (LREs) where there is a LIP model
and showed that it is equivalent to finite-time excitation (FE) or
interval excitation (IE) conditions. The IE/FE condition is known
to be strictly weaker than the persistence of excitation (PE) condition.
Therefore, adaptive methods such as CL (and other methods discussed
in the Introduction) that achieve parameter identification under only
the IE/FE conditions can be interpreted to leverage only the identifiability
of the parameters. In this context, the conditions in Lemma \ref{lem:identifiability}
can be viewed as a generalization of the identifiability conditions
in \cite{Wang.Ortega.ea2021} to NREs where there is an NIP model.
This observation raises the question whether an adaptive estimation
technique can be developed to identify $\theta^{*}$ under the sufficient
identifiability conditions stated in Lemma \ref{lem:identifiability},
as opposed to the more stringent PE condition. Our paper answers this
question with an affirmative by developing a CL-based adaptive estimator
for $\theta^{*}$.%
\begin{comment}
This seems fine
\end{comment}

Before providing the CL-based estimator development, we provide a
brief overview of the gradient-based estimators for DNNs for better
context. For the loss function in (\ref{eq:Loss Function}), the corresponding
loss density is given by $\mathfrak{L}\left(\hat{\theta}\right)=\left\Vert f\left(x\right)-\Phi\left(x,\hat{\theta}\right)\right\Vert ^{2}+\sigma\left\Vert \hat{\theta}\right\Vert ^{2}$.
Then the gradient-based estimator is derived by formulating the negative
gradient flow of $\mathfrak{L}\left(\hat{\theta}\right)$ given by
\begin{eqnarray}
\dot{\hat{\theta}} & = & -\Gamma\nabla_{\hat{\theta}}\mathfrak{L}\left(\hat{\theta}\right)\nonumber \\
 & = & -\Gamma\sigma\hat{\theta}-\Gamma\Phi^{\prime\top}\left(x,\hat{\theta}\right)\tilde{y},\label{eq:Gradient Estimator}
\end{eqnarray}
where $\Gamma\in\mathbb{R}^{\rho\times\rho}$ denotes the adaptation
gain, and 
\begin{equation}
\tilde{y}\triangleq y\left(t\right)-\Phi\left(x,\hat{\theta}\right)\label{eq:prediction_error}
\end{equation}
denotes the regression error. Using a similar approach as in \cite{Patil.Griffis.ea2023}
and \cite{Sweatland.Patil.ea2025}, $\hat{\theta}$ converges to a
neighborhood of $\theta^{*}$ provided the PE condition is satisfied.
The PE condition requires that there exist some $T\in\mathbb{R}_{>0}$
such that $\int_{\underline{t}}^{\underline{t}+T}\Phi^{\prime\top}\left(x(\tau),\hat{\theta}(\tau)\right)\Phi^{\prime}\left(x(\tau),\hat{\theta}(\tau)\right)d\tau\succ0$
for all $\underline{t}\in\mathbb{R}_{\geq0}$, which is restrictive
because the positive definiteness has to hold on a moving time window
for all time. In contrast, Lemma \ref{lem:identifiability} suggests
that $\theta^{*}$ is identifiable if $t\to x(t)$ traverses through
points $x_{1},\ldots,x_{\rho}\in\Omega$ such that conditions ii)
or iii) of Lemma \ref{lem:identifiability} is satisfied. Specifically,
if $t\to x(t)$ is shaped to traverse through such points $x_{1},\ldots,x_{\rho}\in\Omega$
in a finite-time interval, it must be possible to develop an adaptive
estimator to identify $\theta^{*}$ without requiring PE. Based on
this insight, we provide CL-based adaptive update laws in the following
section.

\subsection{Concurrent Learning Algorithm for Continuous-Time Regression Problems}

Consider the NRE in (\ref{eq:mapping f}). Using (\ref{eq:GeneralDNN}),
Remark \ref{rem:Universal Function Approximation}, and (\ref{eq:UFA-1}),
the regression error in (\ref{eq:prediction_error}) can be represented
as
\begin{equation}
\tilde{y}(x)=\Phi(x,\theta^{*})+\varepsilon(x)+\delta(t)-\Phi(x,\hat{\theta}).\label{eq:y_tilde}
\end{equation}
Therefore, the parameter identification objective is to minimize the
parameter estimation error, defined as
\begin{equation}
\tilde{\theta}\triangleq\theta^{*}-\hat{\theta},\label{eq:theta_tilde}
\end{equation}
where $\theta^{*}$ is defined in (\ref{eq:theta star}) and $\hat{\theta}$
represents the parameter estimates. Using the definitions of $\Phi(x,\theta^{*})$
and $\Phi(x,\hat{\theta})$, a first-order Taylor series approximation-based
model of the estimation error is used to obtain \cite{Patil.Le.ea2022}
\begin{equation}
\Phi(x,\theta^{*})-\Phi(x,\hat{\theta})=\Phi^{\prime}(x,\hat{\theta})\tilde{\theta}+R(x,\tilde{\theta}),\label{eq:Taylor-Series}
\end{equation}
where $R(x,\tilde{\theta})$ represents the Lagrange remainder term. 

\subsubsection{Concurrent Learning (CL) Adaptation Laws for Nonlinear Regression}

Two Lyapunov-based adaptation laws which harness different properties
of DNNs are developed to achieve the parameter identification objective.
The first adaptation law is formulated by drawing on established methodologies
in CL using the known input and estimated representation of the system
dynamics, where the estimated representation is formulated using the
output of the DNN. The discrepancy between the known input and the
reconstructed version are then stored in a history stack which provides
a richer data-set that incorporates previous information into the
update law. The second adaptation law uses similar motivation for
the history stack construction; however, it directly incorporates
the Jacobian of the DNN into the update process, leveraging the DNN's
internal dynamics, instead of the output, in the history stack construction.

\paragraph{Weight Adaptation Law Algorithm 1}

The first adaptation law is developed using traditional CL techniques
which allows the adaptation to be guided by the output of the DNN.
The Jacobian of the DNN is denoted $\Phi^{\prime}(x,\hat{\theta})$
and the Jacobian of the DNN at a specified point $x_{i}$ and the
current state estimates is denoted $\Phi^{\prime}(x_{i},\hat{\theta})$.
The core idea is to leverage the recorded input-output data to refine
the parameter estimates. The implementable form of the CL-DNN update
law is designed as 
\begin{equation}
\dot{\hat{\theta}}=\text{proj}\left(\gamma_{1}\sum_{i=1}^{N}\Phi^{\prime\top}(x_{i},\hat{\theta})\left(y_{i}-\hat{y}_{i}\right)-\gamma_{2}\hat{\theta}\right),\label{eq:NLR_ImplementableUpdate1}
\end{equation}
where $\gamma_{1},\gamma_{2}\in\mathbb{R}^{\rho\times\rho}$ denote
positive-definite user-selected adaptation gain matrices and the projection
operator is a smooth projection operator defined in \cite[Appendix E]{Krstic.Kanellakopoulos.ea1995}
which ensures that the adaptive estimate remains within the bounded
parameter search space $\Theta$. An analytical form of the implementable
weight update law can be obtained using (\ref{eq:prediction_error}),
(\ref{eq:y_tilde}), and (\ref{eq:Taylor-Series}) as
\begin{align}
\dot{\hat{\theta}} & =\text{proj}\Bigl(\gamma_{1}\sum_{i=1}^{N}\Phi^{\prime\top}(x_{i},\hat{\theta})\left(\Phi^{\prime}(x_{i},\hat{\theta})\tilde{\theta}\right)-\gamma_{2}\hat{\theta}\nonumber \\
 & \quad+\gamma_{1}\sum_{i=1}^{N}\Phi^{\prime\top}(x_{i},\hat{\theta})\left(R(x_{i},\tilde{\theta})+\varepsilon(x_{i})+\delta(t)\right)\Bigr).\label{eq:NLR_AnalyticalUpdate1}
\end{align}

\paragraph{Weight Adaptation Law Algorithm 2}

The second weight adaptation law is developed based on the desire
to leverage the internal dynamics of the DNN. The implementable form
of the second CL-DNN update law is designed as 
\begin{equation}
\dot{\hat{\theta}}=\text{proj}\left(\gamma_{1}\sum_{i=1}^{N}\Phi^{\prime\top}(x_{i},\hat{\theta})\left(y_{i}+\Phi^{\prime}(x_{i},\hat{\theta})\hat{\theta}\right)-\gamma_{2}\hat{\theta}\right).\label{eq:NLR_ImplementableUpdate2}
\end{equation}
The analytical form of the weight adaptation law can be obtained by
substituting (\ref{eq:mapping f}), (\ref{eq:UFA-1}), and (\ref{eq:Taylor-Series})
into (\ref{eq:NLR_ImplementableUpdate2}) which yields 
\begin{align}
\dot{\hat{\theta}} & =\text{proj}\Bigl(-\gamma_{2}\hat{\theta}+\gamma_{1}\sum_{i=1}^{N}\Phi^{\prime\top}(x_{i},\hat{\theta})\left(\Phi(x_{i},\hat{\theta})+\Phi^{\prime}(x_{i},\hat{\theta})\tilde{\theta}\right.\nonumber \\
 & \quad\left.+R(x_{i},\tilde{\theta}_{i})+\varepsilon(x_{i})+\delta(t)-\Phi^{\prime}(x_{i},\hat{\theta})\hat{\theta}\right)\Bigr).\label{eq:NLR_AnalyticalUpdate2.1}
\end{align}
Using a first-order Taylor series approximation-based model between
the DNN estimate and the DNN with a $0_{\rho}$ weight vector yields
\begin{equation}
-\Phi(x_{i},\hat{\theta})=-\Phi(x_{i},0_{\rho})-\Phi^{\prime}(x_{i},\hat{\theta})\hat{\theta}+R(x_{i},\hat{\theta}_{i}).\label{eq:0Taylor_series}
\end{equation}
Based on the definition in (\ref{eq:Phij_DNN}), the output of the
DNN with all the weights and biases prescribed as $0_{\rho}$ is $\Phi(x_{i},0_{\rho})=0_{n}$.
Using (\ref{eq:0Taylor_series}), (\ref{eq:Analytical_Update2.1}),
and canceling cross terms yields
\begin{align}
\dot{\hat{\theta}} & =\text{proj}\Bigl(\gamma_{1}\sum_{i=1}^{N}\Phi^{\prime\top}(x_{i},\hat{\theta})\left(R(x_{i},\tilde{\theta}_{i})-R(x_{i},\hat{\theta}_{i})+\varepsilon(x_{i})\right)\nonumber \\
 & \quad+\delta(t)+\gamma_{1}\sum_{i=1}^{N}\Phi^{\prime\top}(x_{i},\hat{\theta})\Phi^{\prime}(x_{i},\hat{\theta})\tilde{\theta}-\gamma_{2}\hat{\theta}\Bigr).\label{eq:NLR_AnalyticalUpdate2.2}
\end{align}
The update laws developed in (\ref{eq:NLR_ImplementableUpdate1})
and (\ref{eq:NLR_ImplementableUpdate2}) are evaluated with their
respective analytical forms in (\ref{eq:NLR_AnalyticalUpdate1}) and
(\ref{eq:NLR_AnalyticalUpdate2.2}) in a Lyapunov-based stability
analysis in the following section.

\subsection{Stability Analysis}

Using \cite{Patil.Fallin.ea2025}, the Lagrange remainder in (\ref{eq:Taylor-Series})
is bounded as 
\begin{equation}
\Vert R(x,\tilde{\theta})\Vert\leq\rho_{0}(\left\Vert x\right\Vert )\left\Vert \tilde{\theta}\right\Vert ^{2},\label{eq:LagrangeRemainderBound}
\end{equation}
where $\rho_{0}(\left\Vert x\right\Vert ):\mathbb{R}_{\geq0}\to\mathbb{R}_{\geq0}$
is a strictly increasing quadratic polynomial of the form $\rho_{0}(\Vert x\Vert)=a_{2}\Vert x\Vert^{2}+a_{1}\Vert x\Vert+a_{0}$
with some constants $a_{2},a_{1},a_{0}\in\mathbb{R}_{>0}.$ Similarly,
the Lagrange remainder for (\ref{eq:0Taylor_series}) can be expressed
as $\left\Vert R(x,\hat{\theta})\right\Vert \leq\rho_{0}\left(\left\Vert x\right\Vert \right)\left\Vert \hat{\theta}\right\Vert ^{2}$
which can be further bounded as 
\begin{equation}
\left\Vert R(x,\hat{\theta})\right\Vert \leq\overline{\theta}^{2}\rho_{0}\left(\left\Vert x\right\Vert \right).\label{eq:0LagrangeRemainderBound}
\end{equation}
 Using\cite{Patil.Fallin.ea2025}, the Jacobian of the DNN can be
bounded as
\begin{align}
\left\Vert \Phi^{\prime}(x,\hat{\theta})\right\Vert  & \leq\rho_{1}\left(\left\Vert x\right\Vert \right),\label{eq:Jacobian_Bound}
\end{align}
 where $\rho_{1}:\mathbb{R}_{\geq0}\to\mathbb{R}_{\geq0}$ denotes
a strictly-increasing function. Due to the fact that $R(x,\tilde{\theta}),$
$R(x,\hat{\theta}),$ $\varepsilon(x)$, and $\delta(t)$, are bounded,
the bounds $\left\Vert \Phi^{\prime\top}(x_{i},\hat{\theta})R(x_{i},\tilde{\theta}))\right\Vert \leq\rho_{a}\left(\left\Vert x_{i}\right\Vert \right)\left\Vert \tilde{\theta}\right\Vert ^{2}$,
$\Bigl\Vert\Phi^{\prime\top}(x_{i},\hat{\theta})$ $\left(\varepsilon(x_{i})+\delta(t)\right)\Bigr\Vert$
$\leq\rho_{b}\left(\left\Vert x_{i}\right\Vert \right)$, and $\left\Vert \Phi^{\prime\top}(x_{i},\hat{\theta})\left(\varepsilon(x_{i})+\delta(t)-R(x_{i},\hat{\theta})\right)\right\Vert \leq\rho_{c}\left(\left\Vert x_{i}\right\Vert \right)$
hold where $\rho_{a},\rho_{b},\rho_{c}$$:\mathbb{R}_{\geq0}\to\mathbb{R}_{\geq0}$
denote strictly-increasing bounding functions.

Let $\lambda_{1}\triangleq\frac{\gamma_{2}}{2}-\gamma_{1}+\gamma_{1}\left(\lambda_{\text{min}}\left\{ \sum_{i=1}^{N}\Phi^{\prime\top}(x_{i},\hat{\theta})\Phi^{\prime}(x_{i},\hat{\theta})\right\} \right)$,
$\lambda_{1d}\in\mathbb{R}_{\geq0}$ be the desired convergence rate,
and constants $\iota_{2},\iota_{2}\in\mathbb{R}_{>0}$ be defined
as $\iota_{1}\triangleq$$\frac{\gamma_{2}}{2}\overline{\theta}^{2}+\frac{\gamma_{1}}{2}\left(\sum_{i=1}^{N}\rho_{b}\left(\left\Vert x_{i}\right\Vert \right)\right)^{2}$,
$\ensuremath{\iota_{2}}=\frac{\gamma_{2}}{2}\overline{\theta}^{2}+\frac{\gamma_{1}}{2}\left(\sum_{i=1}^{N}\rho_{c}\left(\left\Vert x_{i}\right\Vert \right)\right)^{2}$.
\begin{assumption}
\label{assum:Assumption-2.FE} There exists $\lambda_{e}>0$ and there
exists $T>0$ for all $t\geq T$ such that $\lambda_{\text{min}}\left\{ \sum_{i=1}^{N}\Phi^{\prime\top}(X_{i},\hat{\theta})\Phi^{\prime}(X_{i},\hat{\theta})\right\} \geq\lambda_{e}$. 
\end{assumption}
\begin{rem}
Since $\lambda_{\text{min}}\left\{ \sum_{i=1}^{N}\Phi^{\prime\top}(X_{i},\hat{\theta})\Phi^{\prime}(X_{i},\hat{\theta})\right\} $
is positive under Assumption \ref{assum:Assumption-2.FE} and is lower-bounded
by $\lambda_{e}$, an increase in $\lambda_{e}$ results in a larger
value for $\lambda_{1}$ being obtained, which implies faster convergence.
If Assumption \ref{assum:Assumption-2.FE} does not hold, the gain
$\gamma_{2}$, which is based on the sigma modification technique
in \cite[Sec. 8.4.1]{Ioannou1996}, helps achieve the boundedness
of $\tilde{\theta}$ in the stability result. However, selecting a
high gain for $\gamma_{2}$ can deteriorate tracking and parameter
estimation performance because it yields a higher value for $\iota_{1}$
and $\iota_{2}$, leading to larger ultimate bounds given by $\sqrt{\frac{\iota_{1}}{\lambda_{1d}}}$
and $\sqrt{\frac{\iota_{2}}{\lambda_{1d}}},$ respectively.
\end{rem}
\begin{rem}
The subsequent theorems, establish ultimate boundedness for the error
signals, rather than the stronger property of uniform ultimate boundedness
(UUB). This distinction, where the characteristics of the ultimate
bound depend on the initial conditions, is a known consequence of
relaxing the PE condition, as observed in the related literature \cite{Ortega.Aranovskiy.2021}
and \cite{Ortega.Romero.2022}. In this result, the dependence on
the initial conditions is primarily due to the history stack which
is populated by the system states which evolve as a function of the
initial condition.
\end{rem}
To facilitate the stability analysis, let the candidate Lyapunov function
$V:\mathbb{R}^{\rho}\to\mathbb{R}_{\geq0}$ be defined as 
\begin{equation}
V(\tilde{\theta})=\frac{1}{2}\tilde{\theta}^{\top}\tilde{\theta}.\label{eq:NLR_V}
\end{equation}
Taking the time derivative of $V(\tilde{\theta})$ yields
\begin{equation}
\dot{V}(\tilde{\theta})=\tilde{\theta}^{\top}\dot{\tilde{\theta}}.\label{eq:NLR_V_Dot}
\end{equation}
Theorem \ref{thm:NLRAlgorithm1} and Theorem \ref{thm:NLRAlgorithm2}
provide convergence guarantees for the parameter estimation errors
for the update laws in (\ref{eq:NLR_AnalyticalUpdate1}) and (\ref{eq:NLR_AnalyticalUpdate2.2}),
respectively. 
\begin{thm}
\label{thm:NLRAlgorithm1}For the parameter identification objective
defined in (\ref{eq:theta_tilde}), the adaptation law developed in
(\ref{eq:NLR_AnalyticalUpdate1}) ensures that the parameter estimation
error $\tilde{\theta}$ is bounded in the sense that $\left\Vert \tilde{\theta}(t)\right\Vert \leq\sqrt{\left\Vert \tilde{\theta}(t_{0})\right\Vert ^{2}e^{-2\lambda_{1d}(t-t_{0})}+\frac{\iota_{1}}{\lambda_{1d}}\left(1-e^{-2\lambda_{1d}t}\right)}$,
for all $t\in\mathbb{R}_{\geq0}$, provided  $\lambda_{1}>\lambda_{1d}+\frac{\gamma_{1}}{2}\rho_{a}^{2}\left(\left\Vert x_{i}\right\Vert \right)\left\Vert \tilde{\theta}(t_{0})\right\Vert ^{2}$.
\end{thm}
\begin{IEEEproof}
Consider the candidate Lyapunov function in (\ref{eq:NLR_V}). From
(\ref{eq:NLR_AnalyticalUpdate1}) and (\ref{eq:NLR_V_Dot}),{\footnotesize{}
\begin{align}
\dot{V}(\tilde{\theta}) & =-\tilde{\theta}^{\top}\Bigl(\gamma_{1}\sum_{i=1}^{N}\Phi^{\prime\top}(x_{i},\hat{\theta})\Phi^{\prime}(x_{i},\hat{\theta})\tilde{\theta}+\gamma_{2}\left(\tilde{\theta}-\theta^{*}\right)\nonumber \\
 & \quad+\gamma_{1}\sum_{i=1}^{N}\Phi^{\prime\top}(x_{i},\hat{\theta})\left(R(x_{i},\tilde{\theta})+\varepsilon(x_{i})+\delta(t)\right)\Bigr).\label{eq:NLR_V_Dot2Algorithm1}
\end{align}
}Upper bounding (\ref{eq:NLR_V_Dot2Algorithm1}) yields 
\begin{align}
\dot{V}(\tilde{\theta}) & \leq-\left(\lambda_{1}-\frac{\gamma_{1}}{2}\sum_{i=1}^{N}\rho_{a}^{2}\left(\left\Vert x_{i}\right\Vert \right)\left\Vert \tilde{\theta}\right\Vert ^{2}\right)\Vert\tilde{\theta}\Vert^{2}\nonumber \\
 & +\frac{\gamma_{1}}{2}\left(\sum_{i=1}^{N}\rho_{b}\left(\left\Vert x_{i}\right\Vert \right)\right)+\frac{\gamma_{2}}{2}\overline{\theta}^{2}.\label{eq:NLR_V_Dot3Algorithm1}
\end{align}
When $\lambda_{1}>\lambda_{1d}+\frac{\gamma_{1}}{2}\rho_{a}^{2}\left(\left\Vert x_{i}\right\Vert \right)\left\Vert \tilde{\theta}(t_{0})\right\Vert ^{2}$,
\begin{equation}
\dot{V}(\tilde{\theta})\leq-\lambda_{1d}\left\Vert \tilde{\theta}\right\Vert ^{2}+\iota_{1}.\label{eq:NLR_FinalVAlgorithm1}
\end{equation}
The inequality in (\ref{eq:NLR_FinalVAlgorithm1}) can be further
bounded as $V(\tilde{\theta}(t))\leq V(\tilde{\theta}(t_{0}))e^{-\lambda_{1d}(t-t_{0})}+\frac{\iota_{1}}{\lambda_{1d}}\left(1-e^{-\textbf{\ensuremath{\lambda_{1d}}}t}\right)$.
Then, \cite[Def. 4.6]{Khalil2002} can be invoked to conclude that
$\tilde{\theta}$ is bounded such that $\left\Vert \tilde{\theta}(t)\right\Vert \leq\sqrt{\left\Vert \tilde{\theta}(t_{0})\right\Vert ^{2}e^{-2\textbf{\ensuremath{\lambda_{1d}}}(t-t_{0})}+\frac{\iota_{1}}{\lambda_{1d}}\left(1-e^{-2\lambda_{1d}t}\right)}$.
Using (\ref{eq:NLR_V}) and (\ref{eq:NLR_FinalVAlgorithm1})\textbf{
}implies $\tilde{\theta}\in\mathcal{L}_{\infty}$. Additionally, due
to the use of the projection operator, $\hat{\theta}\in\mathcal{L}_{\infty}$. 
\end{IEEEproof}
\begin{thm}
\label{thm:NLRAlgorithm2}For the parameter identification objective
defined in (\ref{eq:theta_tilde}), the adaptation law developed in
(\ref{eq:NLR_AnalyticalUpdate1}) ensures that the parameter estimation
error $\tilde{\theta}$ is bounded in the sense that\textbf{ }$\left\Vert \tilde{\theta}(t)\right\Vert \leq\sqrt{\left\Vert \tilde{\theta}(t_{0})\right\Vert ^{2}e^{-2\textbf{\ensuremath{\lambda_{1d}}}(t-t_{0})}+\frac{\iota_{2}}{\lambda_{1d}}\left(1-e^{-2\lambda_{1d}t}\right)}$\textbf{
}, for all $t\in\mathbb{R}_{\geq0}$ provided  $\lambda_{1}>\lambda_{1d}+\frac{\gamma_{1}}{2}\rho_{a}^{2}\left(\left\Vert x_{i}\right\Vert \right)\left\Vert \tilde{\theta}(t_{0})\right\Vert ^{2}$.
\end{thm}
\begin{IEEEproof}
Consider the candidate Lyapunov function in (\ref{eq:NLR_V}). From
(\ref{eq:NLR_V_Dot}) and (\ref{eq:NLR_AnalyticalUpdate2.2}),
\begin{align}
\dot{V}(\tilde{\theta}) & =-\tilde{\theta}^{\top}\Bigl(\gamma_{2}\tilde{\theta}-\gamma_{2}\theta^{*}\nonumber \\
 & \quad+\gamma_{1}\sum_{i=1}^{N}\Phi^{\prime\top}(x_{i},\hat{\theta})\left(\Phi^{\prime}(x_{i},\hat{\theta})\tilde{\theta}+\varepsilon(x_{i})\right.\nonumber \\
 & \quad\left.R(x_{i},\tilde{\theta})-R(x_{i},\hat{\theta})+\delta(t)\right)\Bigr),\label{eq:NLR_V_Dot2Algorithm2}
\end{align}
 and $\dot{V}(\tilde{\theta})$ can be upper bounded as
\begin{equation}
\dot{V}(\tilde{\theta})\leq-\left(\lambda_{1}-\frac{\gamma_{1}}{2}\sum_{i=1}^{N}\rho_{a}^{2}\left(\left\Vert x_{i}\right\Vert \right)\left\Vert \tilde{\theta}\right\Vert ^{2}\right)\left\Vert \tilde{\theta}\right\Vert ^{2}+\iota_{2}.\label{eq:NLR_FinalVAlgorithm2}
\end{equation}
When $\lambda_{1}>\lambda_{1d}+\frac{\gamma_{1}}{2}\sum_{i=1}^{N}\rho_{a}^{2}\left(\left\Vert x_{i}\right\Vert \right)\left\Vert \tilde{\theta}(t_{0})\right\Vert ^{2}$,
\[
\dot{V}(\tilde{\theta})\leq-\lambda_{1d}\left\Vert \tilde{\theta}\right\Vert ^{2}+\iota_{2}.
\]
As a result, $V(\tilde{\theta})$ can be further bounded as $V(\tilde{\theta}(t))\leq V(\tilde{\theta}(t_{0}))e^{-\textbf{\ensuremath{\lambda_{1d}}}(t-t_{0})}+\frac{\iota_{2}}{\lambda_{1d}}\left(1-e^{-\textbf{\ensuremath{\lambda_{1d}}}t}\right)$.
Then, \cite[Def. 4.6]{Khalil2002} can be invoked to conclude that
$\tilde{\theta}$ is bounded such that $\left\Vert \tilde{\theta}(t)\right\Vert \leq\sqrt{\left\Vert \tilde{\theta}(t_{0})\right\Vert ^{2}e^{-2\textbf{\ensuremath{\lambda_{1d}}}(t-t_{0})}+\frac{\iota_{2}}{\lambda_{1d}}\left(1-e^{-2\lambda_{1d}t}\right)}$.
Using (\ref{eq:NLR_V}) and (\ref{eq:NLR_FinalVAlgorithm2})\textbf{
}implies $\tilde{\theta}\in\mathcal{L}_{\infty}$, and by use of the
projection operator, $\hat{\theta}\in\mathcal{L}_{\infty}$.
\end{IEEEproof}

\section{Unknown System Dynamics and Control Design\label{sec:Unknown-System-Dynamics}}

Consider a control-affine nonlinear dynamic system modeled as
\begin{equation}
\ddot{x}=f(x,\dot{x})+u,\label{eq:dynamics}
\end{equation}
where $x,\dot{x}\in\mathbb{R}^{n}$ denotes the measurable generalized
position and velocity, $f:\mathbb{R}^{n}\to\mathbb{R}^{n}$ denotes
an unknown continuously differentiable function, and $u\in\mathbb{R}^{n}$
denotes a control input. The tracking control objective is to simultaneously
track a user-defined reference trajectory $x_{d}\in\mathbb{R}^{n}$
and learn the unknown function $f(x,\dot{x})$ online using a DNN
beginning at the initial time denoted $t_{0}$. Therefore, an unknown
function $f(x,\dot{x})$ can be modeled as
\begin{equation}
f(x,\dot{x})=\Phi(X,\theta^{*})+\varepsilon(X),\label{eq:UFA}
\end{equation}
where the ideal DNN estimate is denoted $\Phi(X,\theta^{*})\in\mathbb{R}^{L_{k+1}}$
where $\theta^{*}$ is defined in (\ref{eq:theta star}), and the
input to the DNN is denoted $X\triangleq\left[x^{\top},\dot{x}^{\top}\right]^{\top}$. 

The reference trajectory is assumed to be sufficiently smooth (i.e.,
$\left\Vert x_{d}\right\Vert \leq\overline{x_{d}}$, $\left\Vert \dot{x}_{d}\right\Vert \leq\overline{\dot{x}_{d}}$,
and $\left\Vert \ddot{x}_{d}\right\Vert \leq\overline{\ddot{x}_{d}}$),
where $\overline{x_{d}},\overline{\dot{x}_{d}},\overline{\ddot{x}_{d}}\in\mathbb{R}_{>0}$
are known constant bounds. To quantify the tracking objective, the
trajectory tracking errors $e,r\in\mathbb{R}^{n}$ are defined as
\begin{align}
e & \triangleq x-x_{d},\label{eq:error}\\
r & \triangleq\dot{e}+\alpha_{1}e.\label{eq:resError}
\end{align}
The structure of (\ref{eq:resError}) is motivated by the subsequent
stability analysis in Theorem \ref{thm:Algorithm1} and Theorem \ref{thm:Algorithm2}
which indicate the boundedness of convergence of $e$ can be determined
from the boundedness and convergence of $r$. As a result, the following
error system development, control design, and stability analysis are
focused on the boundedness and convergence of $r$. Taking the time
derivative of (\ref{eq:resError}) and applying (\ref{eq:dynamics})
yields the open-loop error system
\begin{equation}
\dot{r}=f(x,\dot{x})+u-\ddot{x}_{d}+\alpha_{1}\dot{e}.\label{eq:OLES1-1}
\end{equation}
The subsequent development considers the concurrent tracking control
and system identification problem. It is desirable that the tracking
objective defined in (\ref{eq:error}) is minimized. The parameter
identification objective is to identify a set of parameters that minimizes
the parameter estimation error defined in (\ref{eq:theta_tilde}). 

\subsection{Closed-loop Error System and Control Law Development\label{subsec:Control-Law-Development}}

Since $f(x,\dot{x})$ in (\ref{eq:OLES1-1}) is unknown, we are motivated
to develop an approximation that can be used as a feedforward control
element. The uncertainty in $f(x,\dot{x})$ is not assumed to be modeled
by uncertain linearly parameterizable terms; hence, approximation
methods such as DNN are motivated. In this section, we design the
controller and closed-loop error system to facilitate the subsequent
development of online Lyapunov-based adaptive update laws for the
Lb-CL-DNN and stability analysis. To this end, we substitute the DNN
representation for $f$ from (\ref{eq:UFA}) into (\ref{eq:OLES1-1})
to yield 

\begin{equation}
\dot{r}=\Phi(X,\theta^{*})+\varepsilon(X)+u-\ddot{x}_{d}+\alpha_{1}\dot{e}.\label{eq:OLES2}
\end{equation}
The control input is designed as
\begin{equation}
u=\ddot{x}_{d}-\Phi(X,\hat{\theta})-k_{1}r-e-\alpha_{1}\dot{e},\label{eq:u}
\end{equation}
where $\hat{\theta}\in\mathbb{R}^{\rho}$ is an adaptive estimate
of the parameter $\theta^{*}$, and $k_{1}\in\mathbb{R}_{>0}$ is
a user-defined control gain. Using the definitions of $\Phi(X,\theta^{*})$
and $\Phi(X,\hat{\theta})$, substituting (\ref{eq:u}) and (\ref{eq:Taylor-Series})
onto (\ref{eq:OLES2}) and canceling cross terms yields
\begin{equation}
\dot{r}=\Phi^{\prime}(X,\hat{\theta})\tilde{\theta}+R(x,\tilde{\theta})+\varepsilon(X)-k_{1}r-e.\label{eq:CLES1}
\end{equation}

\subsection{Dynamic State-Derivative Observer\label{subsec:Dynamic-State-Observer}}

Motivated by the desire to incorporate previous state information
into the update law in a CL style, the update law is augmented with
a history stack containing the error between the calculated control
input and a reconstructed version of the control input. To reconstruct
(\ref{eq:dynamics}), an observer is developed to estimate the unmeasurable
state $\ddot{x}$. The dynamic state-derivative observer is designed
as
\begin{align}
\dot{\hat{r}} & =\hat{\Delta}-\ddot{x}_{d}+\alpha_{1}\dot{e}+\alpha_{2}\tilde{r},\label{eq:r_obs_update}\\
\dot{\hat{\Delta}} & =\tilde{r}+k_{\Delta}\tilde{\Delta}-\dot{u},\label{eq:acc_obs_update}
\end{align}
where $\hat{r},\hat{\Delta}\in\mathbb{R}^{n}$ denote the observer
estimates for $r$ and $\ddot{x}$, respectively, $\tilde{r},\tilde{\Delta}\in\mathbb{R}^{n}$
denote the observer errors, $\tilde{r}=r-\hat{r}$ and $\tilde{\Delta}=\ddot{x}-\hat{\Delta}$,
respectively $\alpha,k_{\Delta}\in\mathbb{R}_{>0}$ denote constant
observer gains, and $\dot{u}$ is the unmeasurable derivative of the
control input. Taking the time derivative of $\tilde{r}$ and $\tilde{\Delta}$
and applying (\ref{eq:r_obs_update}) and (\ref{eq:acc_obs_update})
yields
\begin{align}
\dot{\tilde{r}} & =\tilde{\Delta}-\alpha_{2}\tilde{r},\label{eq:tilde_dot_r}\\
\dot{\tilde{\Delta}} & =\dot{f}(x,\dot{x})+\dot{u}-\tilde{r}-k_{\Delta}\tilde{\Delta}-\dot{u},\label{eq:tilde_dot_Delta}
\end{align}
where $\dot{f}(x,\dot{x})\triangleq\frac{d}{dt}f(x,\dot{x})=\frac{\partial f}{\partial x}\dot{x}+\frac{\partial f}{\partial\dot{x}}\ddot{x}$
and $\dot{u}\triangleq\frac{d}{dt}u$. The observer error $\tilde{r}$
is known because $r$ and $\hat{r}$ are known. Because $\tilde{\Delta}=\dot{\tilde{r}}+\alpha_{2}\tilde{r}$,
the implementable form of (\ref{eq:acc_obs_update}) can be obtained
by integrating on both sides and using the relation $\int_{t_{0}}^{t}\dot{\tilde{r}}=\tilde{r}(t)-\tilde{r}(t_{0})$,
which yields $\hat{\Delta}(t)=\hat{\Delta}(t_{0})+k_{\Delta}(\tilde{r}(t)-\tilde{r}(t_{0}))-\left(u(t)-u(t_{0})\right)+\int_{t_{0}}^{t}(k_{\Delta}\alpha_{2}+1)\tilde{r}(\tau)d\tau$.

Let the concatenated error vectors $z\in\mathbb{R}^{2n}$ and $\zeta\in\mathbb{R}^{\psi}$
be defined as $z\triangleq\left[\tilde{r}^{\top},\tilde{\Delta}^{\top}\right]^{\top}$
and $\zeta\triangleq\left[r^{\top},e^{\top},\tilde{\theta}^{\top}\right]^{\top}$,
respectively, where $\psi\triangleq2n+\rho$. Additionally, let the
open and connected set $\mathcal{D}\subset\mathbb{R}^{\psi}$ be defined
as $\mathcal{D}\triangleq\left\{ \zeta\in\rr^{\psi}:\left\Vert \zeta\right\Vert <\chi\right\} $,
where $\chi\in\mathbb{R}_{>0}$ denotes a subsequently defined known
upper bound. The following lemma establishes a bound on $\dot{f}(x,\dot{x})$
when $\zeta\in\mathcal{D}$ to facilitate the convergence analysis
for the observer. The subsequent analysis in the main theorems will
then provide additional conditions under which the trajectories $\zeta(t)$
stay within $\mathcal{D}$ for all time $[t_{0},\infty)$ using the
combined controller-observer adaptation laws.
\begin{lem}
\label{lem:fdot bound} For all $\zeta\in\mathcal{D}$, there exists
a constant $\delta_{f}\in\mathbb{R}_{>0}$ such that the bound $\left\Vert \dot{f}(x,\dot{x})\right\Vert \leq\delta_{f}$
holds.
\end{lem}
\begin{IEEEproof}
For all $\zeta\in\mathcal{D}$, $\left\Vert \zeta\right\Vert <\chi$,
and hence $\left\Vert r\right\Vert ,\left\Vert e\right\Vert ,\left\Vert \tilde{\theta}\right\Vert <\chi$.
Since $\left\Vert x\right\Vert =\left\Vert e+x_{d}\right\Vert \leq\left\Vert e\right\Vert +\left\Vert x_{d}\right\Vert \leq\left\Vert e\right\Vert +\overline{x_{d}}$
and $\left\Vert \dot{x}\right\Vert =\left\Vert r-\alpha_{1}e+\dot{x}_{d}\right\Vert \leq\left\Vert r\right\Vert +\alpha_{1}\left\Vert e\right\Vert +\overline{\dot{x}_{d}}$,
the bounds $\left\Vert x\right\Vert \leq\chi+\overline{x_{d}}$ and
$\left\Vert \dot{x}\right\Vert \leq\left(\alpha_{1}+1\right)\chi+\overline{\dot{x}_{d}}$
hold for all $\zeta\in\mathcal{D}$. Similarly, $\left\Vert \hat{\theta}\right\Vert \leq\left\Vert \theta^{*}\right\Vert +\left\Vert \tilde{\theta}\right\Vert \leq\bar{\theta}+\chi$
for all $\zeta\in\mathcal{D}$. Moreover, using (\ref{eq:dynamics})
and (\ref{eq:u}), $\ddot{x}=f(x,\dot{x})-\Phi(X,\hat{\theta})-k_{1}r-e-\alpha_{1}\dot{e}+\ddot{x}_{d}$.
The terms $f(x,\dot{x})$ and $\Phi(X,\hat{\theta})$ can be bounded
by a constant due to continuity of $f$ and $\Phi$ and boundedness
of $x,\ \dot{x}$, and $\hat{\theta}$, for all $X\in\mathcal{D}$,
because $x,\ \dot{x}$ are bounded and using the definitions in (\ref{eq:error})
and (\ref{eq:resError}), $r,\ e$ and $\dot{e}$ are bounded for
all $X\in\mathcal{D}$ and $\ddot{x}_{d}$ is bounded by design. Thus,
there exists a constant $\overline{\ddot{x}}\in\mathbb{R}_{>0}$ such
that $\left\Vert \ddot{x}\right\Vert \leq\overline{\ddot{x}}$ for
all $\zeta\in\mathcal{D}$. Since $f(x,\dot{x})$ is continuously
differentiable, there exist constants $\varrho_{1},\varrho_{2}\in\mathbb{R}_{>0}$
such that $\left\Vert \frac{\partial f}{\partial x}\right\Vert \leq\varrho_{1}$
and $\left\Vert \frac{\partial f}{\partial\dot{x}}\right\Vert \leq\varrho_{2}$
for all $\zeta\in\mathcal{D}$. Therefore, $\left\Vert \dot{f}(x,\dot{x})\right\Vert \leq\left\Vert \frac{\partial f}{\partial x}\right\Vert \left\Vert \dot{x}\right\Vert +\left\Vert \frac{\partial f}{\partial\dot{x}}\right\Vert \left\Vert \ddot{x}\right\Vert $
and $\left\Vert \ddot{x}\right\Vert \leq\varrho_{1}\left(\left(\alpha_{1}+1\right)\chi+\overline{\dot{x}_{d}}\right)+\varrho_{2}\overline{\ddot{x}}$.
Therefore, selecting $\delta_{f}\triangleq\varrho_{1}\left(\left(\alpha_{1}+1\right)\chi+\overline{\dot{x}_{d}}\right)+\varrho_{2}\overline{\ddot{x}}$
yields $\left\Vert \dot{f}(x,\dot{x})\right\Vert \leq\delta_{f}$
for all $\zeta\in\mathcal{D}$. 
\end{IEEEproof}
Let $\Lambda_{1}\triangleq\text{min}\left\{ k_{\Delta},2\alpha_{2}\right\} $.
The following lemma establishes the convergence properties of the
observer error system in (\ref{eq:tilde_dot_r}) and (\ref{eq:tilde_dot_Delta}).
\begin{lem}
\label{lem:ObserverLemma} Consider the observer given by (\ref{eq:r_obs_update})
and (\ref{eq:acc_obs_update}). The observer error is bounded in the
sense that $\left\Vert z(t)\right\Vert \leq\sqrt{\left\Vert z(t_{0})\right\Vert ^{2}e^{-\Lambda_{1}(t-t_{0})}+\frac{\delta_{f}^{2}}{\Lambda_{1}}\left(1-e^{-\Lambda_{1}t}\right)}$
for all $t\in\mathbb{R}_{\geq0}$, provided $\zeta\in\mathcal{D}$.
Moreover, the observer can achieve a prescribed accuracy $\delta_{\Delta}\in\mathbb{R}_{>0}$
with the settling time $t_{\Delta}\triangleq t_{0}+\frac{1}{\Lambda_{1}}\ln\left(\frac{k_{\Delta}\Lambda_{1}\left\Vert z(t_{0})\right\Vert ^{2}-\delta_{f}^{2}}{k_{\Delta}\Lambda_{1}\delta_{\Delta}^{2}-\delta_{f}^{2}}\right)$,
provided the feasibility gain condition $k_{\Delta}\Lambda_{1}>\frac{\delta_{f}^{2}}{\delta_{\Delta}^{2}}$
is satisfied.
\end{lem}
\begin{IEEEproof}
Consider the candidate Lyapunov function $\mathcal{V}_{\Delta}(z)\triangleq\frac{1}{2}\tilde{\Delta}^{\top}\tilde{\Delta}+\frac{1}{2}\tilde{r}^{\top}\tilde{r}$.
Taking the derivative, using (\ref{eq:tilde_dot_r}), (\ref{eq:tilde_dot_Delta}),
and canceling cross terms yields $\dot{\mathcal{V}}_{\Delta}(z)\leq-\tilde{\Delta}^{\top}k_{\Delta}\tilde{\Delta}-\tilde{r}^{\top}\alpha_{2}\tilde{r}+\tilde{\Delta}^{\top}\dot{f}(x,\dot{x})$.
Using Young's inequality and Lemma \ref{lem:fdot bound}, $\dot{\mathcal{V}}_{\Delta}(z)$
can be further upper-bounded as $\dot{\mathcal{V}}_{\Delta}\leq-\Lambda_{1}\mathcal{V}_{\Delta}+\frac{\delta_{f}^{2}}{2k_{\Delta}}$
provided $\zeta\in\mathcal{D}.$ Therefore, $\mathcal{V}_{\Delta}(z(t))\leq\mathcal{V}_{\Delta}(z(t_{0}))e^{-\Lambda_{1}(t-t_{0})}+\frac{\delta_{f}^{2}}{2k_{\Delta}\Lambda_{1}}\left(1-e^{-\Lambda_{1}t}\right)$
and $\left\Vert z\right\Vert \leq\sqrt{\left\Vert z(t_{0})\right\Vert ^{2}e^{-\Lambda_{1}(t-t_{0})}+\frac{\delta_{f}^{2}}{k_{\Delta}\Lambda_{1}}\left(1-e^{-\Lambda_{1}t}\right)}$,
provided $\zeta\in\mathcal{D}.$ For the prescribed accuracy $\delta_{\Delta}$,
using the differential inequality, the settling time $t_{\Delta}=t_{0}+\frac{1}{\Lambda_{1}}\ln\left(\frac{k_{\Delta}\Lambda_{1}\Vert z(t_{0})\Vert^{2}-\delta_{f}^{2}}{k_{\Delta}\Lambda_{1}\delta_{\Delta}^{2}-\delta_{f}^{2}}\right)$
is obtained after imposing $\delta_{\Delta}\geq\sqrt{\left\Vert z(t_{0})\right\Vert ^{2}e^{-\Lambda_{1}(t-t_{0})}+\frac{\delta_{f}^{2}}{k_{\Delta}\Lambda_{1}}\left(1-e^{-\Lambda_{1}t}\right)}$,
provided $\zeta\in\mathcal{D}.$ For the settling time to be feasible,
the argument of the natural logarithm needs to be positive; imposing
this condition yields the feasibility gain condition $k_{\Delta}\Lambda_{1}>\frac{\delta_{f}^{2}}{\delta_{\Delta}^{2}}$.
\end{IEEEproof}
Using Lemma \ref{lem:ObserverLemma}, the observer errors, $\tilde{\Delta}$
and $\tilde{r}$ will converge to the prescribed ultimate bound $\delta_{\Delta}$
after the time $t_{\Delta}$ has passed. 

\section{CL Adaptation Laws For Adaptive Control\label{sec:Concurrent-Learning-(CL)UpdateLaws}}

\subsection{Weight Adaptation Law Algorithm 1}

The first adaptation law is developed using traditional CL techniques
guided by the output of the DNN. The shorthand notation $\Phi^{\prime}(X,\hat{\theta})$
in (\ref{eq:Implementable_Update}) denotes the Jacobian of the DNN
at the current state and weight estimates, and $\Phi^{\prime\top}(X_{i},\hat{\theta})$
represents the Jacobian of the DNN using the current weight estimates
and the previous state $X_{i}$, where $X_{i}\triangleq X(t_{i})$
and $t_{i}\in\left[t_{\Delta},t\right]$, represents states within
this time interval. While some concurrent learning methods initialize
the history stack at $t=0$, in this approach, the history stack is
only constructed after the settling time ($t_{\Delta}$ defined in
Lemma \ref{lem:ObserverLemma}) has been reached. For the history
stack to be accurate, this data is gathered after the observer errors
have reached their ultimate bound. The implementable form of the CL-DNN
update law is designed as
\begin{align}
\dot{\hat{\theta}} & =\text{proj}\Bigl(\Gamma\left(\Phi^{\prime\top}(X,\hat{\theta})r-\gamma_{1}\sum_{i=1}^{N}\Phi^{\prime\top}(X_{i},\hat{\theta})\left(u_{i}-\hat{u}_{i}\right)\right.\nonumber \\
 & \;-\left.\gamma_{2}\hat{\theta}\right)\Bigr),\label{eq:Implementable_Update}
\end{align}
where $\gamma_{1},\gamma_{2}\in\mathbb{R}_{>0}$ denote user-selected
adaptation gains, and $\Gamma\in\mathbb{R}^{\rho\times\rho}$ denotes
a positive-definite time-varying least squares adaptation gain matrix
{\footnotesize{}
\begin{equation}
\frac{d}{dt}\Gamma^{-1}=\begin{cases}
-\beta\Gamma^{-1}+\gamma_{1}\left(\sum_{i=1}^{N}\Phi^{\prime\top}(X_{i},\hat{\theta})\Phi^{\prime}(X_{i},\hat{\theta})\right),\\
\quad\quad\quad\text{if }\lambda_{\Gamma,\text{min}}<\lambda_{\text{min}}\left(\Gamma\right)\text{ and }\text{\ensuremath{\lambda_{\text{max}}}\ensuremath{\left(\Gamma\right)}}<\lambda_{\Gamma,\text{max}}\\
0,\quad\quad\text{otherwise}
\end{cases}\label{eq:AdaptiveGamma}
\end{equation}
}where $\beta$ represents a user selected forgetting factor $\beta:\mathbb{R}_{\geq0}\to\mathbb{R}_{\geq0}$,
and $\lambda_{\Gamma,\text{min}},$ $\lambda_{\Gamma,\text{max }}$
are user-selected bounds for the minimum and maximum eigenvalues of
$\Gamma$, respectively. The adaptation gain in (\ref{eq:AdaptiveGamma})
is initialized to be PD and it can be shown that $\Gamma(t)$ remains
PD for all $t\in\mathbb{R}_{\geq0}$ \cite{Slotine1989}. A reconstructed
estimate of the calculated control input can be determined as
\begin{equation}
\hat{u}_{i}=\hat{\Delta}_{i}-\Phi(X_{i},\hat{\theta}).\label{eq:est_control_input}
\end{equation}
An analytical form of the implementable weight adaptation law in (\ref{eq:Implementable_Update})
can be obtained from (\ref{eq:dynamics}), (\ref{eq:UFA}), (\ref{eq:Taylor-Series}),
(\ref{eq:Implementable_Update}), and (\ref{eq:est_control_input})
as{\small{}
\begin{align}
\dot{\hat{\theta}} & =\text{proj}\Bigl(\Gamma\left(\Phi^{\prime\top}(X,\hat{\theta})r-\gamma_{2}\hat{\theta}+\gamma_{1}\sum_{i=1}^{N}\Phi^{\prime\top}(X_{i},\hat{\theta})\Phi^{\prime}(X_{i},\hat{\theta})\tilde{\theta}\right.\nonumber \\
 & \;+\left.\gamma_{1}\sum_{i=1}^{N}\Phi^{\prime\top}(X_{i},\hat{\theta})\left(R(x,\tilde{\theta})+\varepsilon(X_{i})-\tilde{\Delta}_{i}\right)\right)\Bigr).\label{eq:analytical_update1}
\end{align}
}{\small\par}

\subsection{Weight Adaptation Law Algorithm 2 }

The second weight adaptation law is developed to achieve parameter
convergence and is based on the desire to leverage the internal dynamics
of the DNN. The implementable form of the second CL-DNN update law
is designed as
\begin{align}
\dot{\hat{\theta}} & =\text{proj}\Bigl(\Gamma\Bigl(\Phi^{\prime\top}(X,\hat{\theta})r-\gamma_{2}\hat{\theta}\nonumber \\
 & \;-\gamma_{1}\sum_{i=1}^{N}\Phi^{\prime\top}(X_{i},\hat{\theta})\left(u_{i}-\hat{\Delta}_{i}+\Phi^{\prime}(X_{i},\hat{\theta})\hat{\theta}\right)\Bigr)\Bigr).\label{eq:Implementable_Update2}
\end{align}
The analytical form of the weight adaptation law can be obtained by
substituting (\ref{eq:dynamics}), (\ref{eq:UFA}), and (\ref{eq:Taylor-Series})
into (\ref{eq:Implementable_Update2}) which yields
\begin{align}
\dot{\hat{\theta}} & =\text{proj}\Bigl(\Gamma\left(\Phi^{\prime\top}(X,\hat{\theta})r-\gamma_{1}\sum_{i=1}^{N}\Phi^{\prime\top}(X_{i},\hat{\theta})\left(\tilde{\Delta}_{i}\right.\right.\nonumber \\
 & \;-\Phi(X_{i},\hat{\theta})-\Phi^{\prime}(X_{i},\hat{\theta})\tilde{\theta}-R(X_{i},\tilde{\theta})+\varepsilon(X_{i})\nonumber \\
 & \;\left.+\Phi^{\prime}(X_{i},\hat{\theta})\hat{\theta}\right)\left.-\gamma_{2}\hat{\theta}\right)\Bigr).\label{eq:Analytical_Update2.1}
\end{align}
Based on the definition in (\ref{eq:Phij_DNN}), $\Phi(X,0_{\rho})=0_{n}$.
Using (\ref{eq:0Taylor_series}), (\ref{eq:Analytical_Update2.1}),
and canceling cross terms yields{\small{}
\begin{align}
\dot{\hat{\theta}} & =\text{proj}\Bigl(\Gamma\left(\Phi^{\prime\top}(X,\hat{\theta})r+\gamma_{1}\sum_{i=1}^{N}\Phi^{\prime\top}(X_{i},\hat{\theta})\Phi^{\prime}(X_{i},\hat{\theta})\tilde{\theta}\right.\nonumber \\
 & \;-\gamma_{1}\sum_{i=1}^{N}\Phi^{\prime\top}(X_{i},\hat{\theta})\left(\tilde{\Delta}_{i}+R(X_{i},\hat{\theta})-R(X_{i},\tilde{\theta})+\varepsilon(X_{i})\right)\nonumber \\
 & \;\left.-\gamma_{2}\hat{\theta}\right)\Bigr).\label{eq:Analytical_Update2.2}
\end{align}
}The update laws developed in (\ref{eq:Implementable_Update}) and
(\ref{eq:Implementable_Update2}) are evaluated with their respective
analytical forms in (\ref{eq:analytical_update1}) and (\ref{eq:Analytical_Update2.2})
and the controller developed in (\ref{eq:u}) in a Lyapunov-based
stability analysis in the following section.

\section{Stability Analysis\label{sec:Stability-AnalysisAdaptive}}

Using\textbf{ }(\ref{eq:LagrangeRemainderBound}), the following
bound can be established
\begin{align}
\left\Vert R(X,\tilde{\theta})\right\Vert  & \leq\rho_{2}\left(\left\Vert \zeta\right\Vert \right)\left\Vert \zeta\right\Vert ^{2}\label{eq:Jacobian_Lagrange_Bounds_Control}
\end{align}
 where $\rho_{2}:\mathbb{R}_{\geq0}\to\mathbb{R}_{\geq0}$ denotes
a strictly-increasing function. Due to the fact that $R(X,\tilde{\theta})$,
$\Phi^{\prime}(X,\hat{\theta})$, and $\varepsilon$ are bounded,
then the following bounds can be established
\begin{align}
\left\Vert \Phi^{\prime\top}(X_{i},\hat{\theta})R(X_{i},\tilde{\theta})\right\Vert  & \leq\rho_{3}\left(\left\Vert \zeta_{i}\right\Vert \right)\left\Vert \zeta\right\Vert ^{2},\nonumber \\
\left\Vert \Phi^{\prime\top}(X_{i},\hat{\theta})\varepsilon(X_{i})\right\Vert  & \leq\rho_{4}\left(\left\Vert X_{i}\right\Vert \right),\nonumber \\
\left.\left\Vert \Phi^{\prime\top}(X_{i},\hat{\theta})\left(\varepsilon(X_{i})+R(X_{i},\hat{\theta})\right)\right.\right\Vert  & \leq\rho_{5}\left(\left\Vert X_{i}\right\Vert \right),\label{eq:HS_rho_bounds}
\end{align}
where $\rho_{3},\rho_{4},\rho_{5}:\mathbb{R}_{\geq0}\to\mathbb{R}_{\geq0}$
denote strictly-increasing functions. Let $\rho_{\Delta}\left(\left\Vert \zeta\right\Vert \right)\geq\frac{1}{2}\rho_{2}^{2}\left(\left\Vert \zeta\right\Vert \right)\left\Vert \zeta\right\Vert ^{2}+\frac{\gamma_{1}}{2}\sum_{i=1}^{N}\rho_{3}^{2}\left(\left\Vert \zeta_{i}\right\Vert \right)\left\Vert \zeta\right\Vert ^{2}$,
$\rho_{\delta}\left(\left\Vert \zeta\right\Vert \right)\geq\frac{1}{2}\rho_{1}^{2}\left(\left\Vert \zeta\right\Vert \right)\left\Vert \zeta\right\Vert ^{2}+\frac{\gamma_{1}}{2}\sum_{i=1}^{N}\rho_{3}^{2}\left(\left\Vert \zeta_{i}\right\Vert \right)\left\Vert \zeta\right\Vert ^{2}$,
where $\rho_{\Delta},\rho_{\delta}:\mathbb{R}_{\geq0}\to\mathbb{R}_{\geq0}$
denote invertible strictly-increasing functions. Since the approximation
capabilities of DNNs holds on a compact domain $\Omega$, the subsequent
stability analysis requires ensuring $X(t)\in\Omega$ for all $t\in[t_{0},\infty)$.
This is achieved by yielding a stability result which constrains $\zeta$
in a compact domain. Therefore, consider the compact domains $\mathcal{D}_{3,4}\triangleq\left\{ \sigma\in\mathbb{R}^{\psi}:\left\Vert \sigma\right\Vert \leq\chi_{3,4}\right\} $
in which $\zeta$ is supposed to lie to develop Theorem \ref{thm:Algorithm1}
and \ref{thm:Algorithm2}, respectively. It follows that if $\left\Vert \zeta\right\Vert \leq\chi_{3,4}$
then $X$ can be bounded as $\left\Vert X\right\Vert \leq\left(\alpha+2\right)\chi_{3,4}+\overline{x_{d}}+\overline{\dot{x}_{d}}$.
Therefore, select $\Omega_{3,4}\triangleq\{\sigma\in\mathbb{R}^{2n}:\left\Vert \sigma\right\Vert \leq\left(\alpha_{1}+2\right)\chi_{3,4}+\overline{x_{d}}+\overline{\dot{x}_{d}}\}$.
Then $\zeta\in\mathcal{D}_{3,4}$ implies $X\in\Omega_{3,4}$. Furthermore,
to ensure that arbitrary initial conditions are always included, the
user-selected constants $\chi_{3}$ and $\chi_{4}$ are selected as
\begin{align}
\chi_{3} & \triangleq\rho_{\Delta}^{-1}\left(\lambda_{3}-\lambda_{3d}\right),\nonumber \\
\chi_{4} & \triangleq\rho_{\delta}^{-1}\left(\lambda_{3}-\lambda_{3d}\right).\label{eq:Chi Definition}
\end{align}
Then it follows that $z(t_{0})\in\mathcal{D}_{3,4}=\left\{ \zeta\in\mathbb{R}^{2n+p}:\left\Vert \zeta\right\Vert \leq\chi\right\} $
is always satisfied. Because the solution $t\mapsto\zeta(t)$ is continuous\footnote{Continuous solutions exists over some time-interval for systems satisfying
Caratheodory existence conditions. According to Caratheodory conditions
for the system $\dot{y}=f(y,t)$, $f$ should be locally bounded,
continuous in $y$ for each fixed $t$, measurable in $t$ for each
fixed $y$ \cite[Ch.2, Theorem 1.1]{Coddington.Levinson1955}. The
dynamics in $\dot{z}$ satisfy the Caratheodory conditions.}, there exists a time-interval $\mathcal{I}_{1,2}\triangleq[t_{0},t_{1})$
such that $\left\Vert \zeta(t)\right\Vert <\mathcal{D}_{3,4}$ for
all $t\in\mathcal{I}_{1,2}$. It follows that $X(t)\in\Omega_{3,4}$
for all $t\in\mathcal{I}_{1,2}$, therefore the universal function
approximation property holds over this time interval. In the subsequent
stability analysis, we analyze the convergence properties of the solutions
and also establish that $\mathcal{I}_{1,2}$ can be extended to $[t_{0},\infty)$.
Let $\textbf{\ensuremath{\lambda_{3}}}\triangleq\text{min}\Bigl\{ k_{1}-1,\alpha_{1},\frac{\gamma_{1}}{2}\left(\text{\ensuremath{\lambda_{\text{min}}}}\left\{ \sum_{i=1}^{N}\Phi^{\prime\top}(X_{i},\hat{\theta})\Phi^{\prime}(X_{i},\hat{\theta})\right\} \right)+\frac{\gamma_{2}}{2}-\frac{\gamma_{1}N\delta_{\Delta}}{2}+\gamma_{1}\Bigr\}$,
$\iota_{3}\triangleq\frac{1}{2}\overline{\varepsilon}^{2}+\frac{\gamma_{2}}{2}\overline{\theta}^{2}+\frac{\gamma_{1}N\delta_{\Delta}}{2}\sum_{i=1}^{N}\rho_{1}\left(\left\Vert X_{i}\right\Vert \right)-\frac{\gamma_{1}}{2}\sum_{i=1}^{N}\left(\rho_{4}\left(\left\Vert X_{i}\right\Vert \right)\right)^{2}$
and $\iota_{4}\triangleq\frac{1}{2}\overline{\varepsilon}^{2}+\frac{\gamma_{1}N\delta_{\Delta}}{2}\sum_{i=1}^{N}\left(\rho_{1}\left(\left\Vert X_{i}\right\Vert \right)\right)^{2}+\frac{\gamma_{1}}{2}\sum_{i=1}^{N}\left(\rho_{5}\left(\left\Vert X_{i}\right\Vert \right)\right)^{2}+\frac{\gamma_{2}}{2}\overline{\theta}^{2}$.
Let the sets $\mathcal{S}\subset\mathcal{D}_{3}$ and $\mathcal{H}\subset\mathcal{D}_{4}$
be the sets of stabilizing initial conditions defined as
\begin{align*}
\mathcal{S} & \triangleq\left\{ \sigma\in\mathbb{R}^{\psi}:\left\Vert \zeta(t_{0})\right\Vert <\sqrt{\frac{\beta_{1}}{\beta_{2}}\rho_{\Delta}^{-1}\left(\lambda_{3}-\lambda_{3d}\right)^{2}-\frac{\iota_{3}}{\lambda_{3d}}}\right\} ,\\
\mathcal{H} & \triangleq\left\{ \sigma\in\mathbb{R}^{\psi}:\left\Vert \zeta(t_{0})\right\Vert <\sqrt{\frac{\beta_{1}}{\beta_{2}}\rho_{\delta}^{-1}\left(\lambda_{3}-\lambda_{3d}\right)^{2}-\frac{\iota_{4}}{\lambda_{3d}}}\right\} ,
\end{align*}
where $\lambda_{3d}\in\mathbb{R}_{\geq0}$ denotes the desired convergence
rate.

To guarantee that the adaptive update laws in Theorem \ref{thm:Algorithm1}
and Theorem \ref{thm:Algorithm2} achieve the parameter identification
objective described in (\ref{eq:theta_tilde}), the history stack
needs to be sufficiently rich, i.e., the system needs to be excited
over a finite duration of time as specified in the following assumption.
\begin{assumption}
\label{assum:FE_Controls} There exists $\lambda_{e}>0$ and there
exists $T>t_{\Delta}$ for all $t\geq T$ such that $\lambda_{\text{min}}\left\{ \sum_{i=1}^{N}\Phi^{\prime\top}(X_{i},\hat{\theta})\Phi^{\prime}(X_{i},\hat{\theta})\right\} \geq\lambda_{e}$.
\end{assumption}
To promote exploration of the parameter space, which is desirable
to satisfy Assumption \ref{assum:Assumption-2.FE}, the weights could
be artificially excited as detailed in the following remark.
\begin{rem}
\label{rem:time varying disurbance}To promote exploration of the
weights of the DNN, a user-selected time-varying bounded dither signal
$d(t):\mathbb{R}_{\geq0}\to\mathbb{R}^{\rho}$ with the bound $\underset{t\geq0}{\sup}$$\left\Vert d(t)\right\Vert \leq\overline{d}$
can be injected into the adaptation laws in (\ref{eq:analytical_update1})
and (\ref{eq:Analytical_Update2.2}). The resulting $\iota_{3}$ and
$\iota_{4}$ would contain an additional term $\frac{\gamma_{3}}{2}\overline{d}^{2}$
where $\gamma_{3}\in\mathbb{R}_{>0}$ is a user-selected gain.
\end{rem}
Consider the Lyapunov candidate function $\mathcal{V}:\mathbb{R}^{\psi}\to\mathbb{R}_{\geq0}$
defined as 
\begin{equation}
\mathcal{V}(\zeta)=\frac{1}{2}r^{\top}r+\frac{1}{2}e^{\top}e+\frac{1}{2}\tilde{\theta}^{\top}\Gamma^{-1}\tilde{\theta},\label{eq:V}
\end{equation}
which satisfies the inequality 
\begin{equation}
\beta_{1}\left\Vert \zeta\right\Vert ^{2}\leq\mathcal{V}(\zeta)\leq\beta_{2}\left\Vert \zeta\right\Vert ^{2}\label{eq:LyapBounds}
\end{equation}
 where $\beta_{1}\triangleq\text{min}\left\{ \frac{1}{2},\frac{1}{2},\frac{1}{2}\frac{1}{\lambda_{\Gamma,\text{max}}}\right\} $
and $\beta_{2}\triangleq\text{max}\left\{ \frac{1}{2},\frac{1}{2},\frac{1}{2}\frac{1}{\lambda_{\Gamma,\text{min}}}\right\} $.
Taking the time derivative of $\mathcal{V}(\zeta)$, and applying
(\ref{eq:resError}) and (\ref{eq:CLES1}) yields
\begin{align}
\dot{\mathcal{V}} & =r^{\top}\left(\Phi^{\prime}(X,\hat{\theta})\tilde{\theta}+R(X,\tilde{\theta})+\varepsilon(X)-k_{1}r-e\right)\nonumber \\
 & \;-e^{\top}\alpha_{1}e+\tilde{\theta}^{\top}\Gamma^{-1}\dot{\tilde{\theta}}+\frac{1}{2}\tilde{\theta}^{\top}\left(\frac{d}{dt}\Gamma^{-1}\right)\tilde{\theta}.\label{eq:V_Dot}
\end{align}
Using (\ref{eq:AdaptiveGamma}), the last term in (\ref{eq:V_Dot})
can be bounded as {\footnotesize{}
\begin{align}
\frac{1}{2}\tilde{\theta}^{\top}\left(\frac{d}{dt}\Gamma^{-1}\right)\tilde{\theta} & \leq\frac{1}{2}\tilde{\theta}^{\top}\gamma_{1}\left(\sum_{i=1}^{N}\Phi^{\prime\top}(X_{i},\hat{\theta})\Phi^{\prime}(X_{i},\hat{\theta})\right)\tilde{\theta}.\label{eq:GammaInequality}
\end{align}
}Theorem \ref{thm:Algorithm1} and Theorem \ref{thm:Algorithm2} provide
convergence guarantees for the tracking and parameter estimation errors
using the update laws in (\ref{eq:analytical_update1}) and (\ref{eq:Analytical_Update2.2}),
respectively.
\begin{thm}
\label{thm:Algorithm1}Let the gain conditions $k_{\Delta}\Lambda_{1}>\frac{\delta_{f}^{2}}{\delta_{\Delta}^{2}}$
and $\lambda_{3d}>0$ be satisfied, and $\left\Vert \zeta(0)\right\Vert \in\mathcal{S}$.
For the dynamical system in (\ref{eq:dynamics}), the controller in
(\ref{eq:u}) and the adaptation law developed in (\ref{eq:analytical_update1})
ensures the concatenated error vector $\zeta$ is bounded in the sense
that $\left\Vert \zeta(t)\right\Vert \Vert\leq\sqrt{\frac{\beta_{2}}{\beta_{1}}\left\Vert \zeta(t_{0})\right\Vert ^{2}e^{-\frac{\lambda_{3d}}{\beta_{2}}(t-t_{0})}+\frac{\beta_{2}\iota_{3}}{\beta_{1}\lambda_{3d}}\left(1-e^{-\frac{\lambda_{3d}}{\beta_{2}}t}\right)}$,
for all $t\in\mathbb{R}_{\geq0}$. 
\end{thm}
\begin{IEEEproof}
Consider the candidate Lyapunov function in (\ref{eq:V}). From (\ref{eq:analytical_update1}),
(\ref{eq:V_Dot}), and using (\ref{eq:GammaInequality}), $\dot{\mathcal{V}}$
can be upper bounded as 
\begin{align}
\dot{\mathcal{V}} & \leq-r^{\top}k_{1}r-e^{\top}\alpha_{1}e+r^{\top}\left(R(X,\tilde{\theta})+\varepsilon(X)\right)\nonumber \\
 & \;-\tilde{\theta}^{\top}\frac{\gamma_{1}}{2}\sum_{i=1}^{N}\Phi^{\prime\top}(X_{i},\hat{\theta})\Phi^{\prime}(X_{i},\hat{\theta})\tilde{\theta}-\tilde{\theta}^{\top}\gamma_{2}\tilde{\theta}\nonumber \\
 & \;-\tilde{\theta}^{\top}\gamma_{1}\sum_{i=1}^{N}\Phi^{\prime\top}(X_{i},\hat{\theta})\left(R(X_{i},\tilde{\theta})+\varepsilon(X_{i})-\tilde{\Delta}_{i}\right)\nonumber \\
 & \;+\tilde{\theta}^{\top}\gamma_{2}\theta^{*}.\label{eq:V_Dot2Algorithm1}
\end{align}
Using Lemma \ref{lem:ObserverLemma}, the bound $\left\Vert \tilde{\Delta}_{i}\right\Vert \leq\delta_{\Delta}$
holds for all $i\in\left\{ 1,\ldots,N\right\} $, provided $\zeta\in\mathcal{D}_{3}$,
the data is collected after the settling time $t_{\Delta}$, and the
feasibility gain condition $k_{\Delta}\Lambda_{1}>\frac{\delta_{f}^{2}}{\delta_{\Delta}^{2}}$
is satisfied. Therefore, using the bounds in (\ref{eq:Jacobian_Lagrange_Bounds_Control})
and (\ref{eq:HS_rho_bounds}), $\dot{\mathcal{V}}$ can be upper bounded
as
\begin{equation}
\dot{\mathcal{V}}\leq-\left(\lambda_{3}-\rho_{\Delta}\left(\left\Vert \zeta\right\Vert \right)\right)\left\Vert \zeta\right\Vert ^{2}+\iota_{3},\label{eq:V_Dot3Algorithm1}
\end{equation}
when $\zeta\in\mathcal{D}_{3}$. Therefore, when $\zeta$ is initialized
such that $\zeta(t_{0})\in\mathcal{S}$, then it follows that $\textbf{\ensuremath{\lambda_{3}}}>\lambda_{3d}+\rho_{\Delta}\left(\sqrt{\frac{\beta_{2}}{\beta_{1}}\left\Vert \zeta(t_{0})\right\Vert ^{2}+\frac{\beta_{2}\iota_{3}}{\beta_{1}\lambda_{3d}}}\right)$.
Recall that, because the solution $t\mapsto\zeta(t)$ is continuous,
$\zeta$ cannot instantaneously escape $\mathcal{S}$ at $t_{0}$,
therefore there exists a time-interval $\mathcal{I}_{1}$ such that
$\zeta(t)\in\mathcal{S}$ for all $t\in\mathcal{I}_{1}$, implying
\begin{equation}
\left\Vert \zeta(t)\right\Vert <\sqrt{\frac{\beta_{2}}{\beta_{1}}\left\Vert \zeta(t_{0})\right\Vert ^{2}+\frac{\beta_{2}\iota_{3}}{\beta_{1}\lambda_{3d}}}\label{eq:z_bound}
\end{equation}
for all $t\in\mathcal{I}_{1}$. Because $\rho_{\Delta}$ is strictly
increasing, $\rho_{\Delta}\left(\left\Vert \zeta(t)\right\Vert \right)<\rho_{\Delta}\left(\sqrt{\frac{\beta_{2}}{\beta_{1}}\left\Vert \zeta(t_{0})\right\Vert ^{2}+\frac{\beta_{2}\iota_{3}}{\beta_{1}\lambda_{3d}}}\right)$
for all $t\in\mathcal{I}_{1}$. As a result,
\begin{equation}
\dot{\mathcal{V}}\leq-\lambda_{3d}\left\Vert \zeta\right\Vert ^{2}+\iota_{3}\label{eq:FinalV_Algorithm1}
\end{equation}
for all $t\in\mathcal{I}_{1}.$ Using (\ref{eq:LyapBounds}), $\dot{\mathcal{V}}$
and solving the differential inequality over $\mathcal{I}_{1}$ yields
 $\mathcal{V}(\zeta(t))\leq\mathcal{V}(\zeta(t_{0}))e^{-\frac{\lambda_{3d}}{\beta_{2}}(t-t_{0})}+\frac{\beta_{2}\iota_{3}}{\lambda_{3d}}\left(1-e^{-\frac{\lambda_{3d}}{\beta_{2}}t}\right)$.
It remains to be shown that $\mathcal{I}_{1}$ can be extended to
$[t_{0},\infty)$. Assume for the sake of contradiction that $\mathcal{I}_{1}$
has to be bounded, i.e., $t_{1}$ has to be finite. Equivalently,
there exists $t_{1}$ for which there does not exist $t_{2}>t_{1}$
such that $\left\Vert \zeta(t)\right\Vert <\sqrt{\frac{\beta_{2}}{\beta_{1}}\left\Vert \zeta(t_{0})\right\Vert ^{2}+\frac{\beta_{2}\iota_{3}}{\beta_{1}\lambda_{3d}}}$
for all $t\in[t_{1},t_{2})$. Substituting $t=t_{1}$ into (\ref{eq:z_bound})
yields $\left\Vert \zeta(t_{1})\right\Vert <\sqrt{\frac{\beta_{2}}{\beta_{1}}\left\Vert \zeta(t_{0})\right\Vert ^{2}+\frac{\beta_{2}\iota_{3}}{\beta_{1}\lambda_{3d}}}$
. Hence, because the solution $t\mapsto\zeta(t)$ is continuous, for
every $t_{1}$ there exists a $t_{2}>t_{1}$ such that $\left\Vert \zeta(t)\right\Vert <\sqrt{\frac{\beta_{2}}{\beta_{1}}\left\Vert \zeta(t_{0})\right\Vert ^{2}+\frac{\beta_{2}\iota_{3}}{\beta_{1}\lambda_{3d}}}$for
all $t\in[t_{1},t_{2}),$ violating the assumption made by contradiction.
Therefore, $\mathcal{I}_{1}$ can be extended to the interval $[t_{0},\infty)$
and $\left\Vert \zeta(t)\right\Vert <\sqrt{\frac{\beta_{2}}{\beta_{1}}\left\Vert \zeta(t_{0})\right\Vert ^{2}+\frac{\beta_{2}\iota_{3}}{\beta_{1}\lambda_{3d}}}$
for all $t\in[t_{0},\infty)$. Then, \cite[Def. 4.6]{Khalil2002}
can be invoked to conclude that $\zeta$ is bounded such that $\left\Vert \zeta(t)\right\Vert \Vert\leq\sqrt{\frac{\beta_{2}}{\beta_{1}}\left\Vert \zeta(t_{0})\right\Vert ^{2}e^{-\frac{\lambda_{3d}}{\beta_{2}}(t-t_{0})}+\frac{\beta_{2}\iota_{3}}{\beta_{1}\lambda_{3d}}\left(1-e^{-\frac{\lambda_{3d}}{\beta_{2}}t}\right)}$.
Hence, solving the differential inequality in (\ref{eq:FinalV_Algorithm1})
and upperbounding yields that if $\zeta(t_{0})\in\mathcal{S}$, then
$\zeta(t)\in\mathcal{S}\subset\mathcal{D}_{3}$ and therefore $X\in\Omega_{3}$
for all $t\geq0$. Using (\ref{eq:V}) and (\ref{eq:FinalV_Algorithm1})
implies $e,r,\tilde{\theta}\in\mathcal{L}_{\infty}.$ The fact that
$x_{d},\dot{x}_{d},\ddot{x}_{d},e,r\in\mathcal{L}_{\infty}$ implies
$x,\dot{x}\in\mathcal{L}_{\infty}$, using this and the fact $\hat{\theta}\in\mathcal{L}_{\infty}$
is bounded by the use of the projection operator implies $u$ is bounded.
\end{IEEEproof}
\begin{thm}
\label{thm:Algorithm2}Let the gain conditions $k_{\Delta}\Lambda_{1}>\frac{\delta_{f}^{2}}{\delta_{\Delta}^{2}}$
and $\lambda_{3d}>0$ be satisfied, and $\Vert\zeta(0)\Vert\in\mathcal{H}$.
For the dynamical system in (\ref{eq:dynamics}), the controller in
(\ref{eq:u}), the adaptation law developed in (\ref{eq:Analytical_Update2.2})
ensures the concatenated error vector $\zeta$ is bounded in the sense
that $\left\Vert \zeta(t)\right\Vert \leq\sqrt{\frac{\beta_{2}}{\beta_{1}}\left\Vert \zeta(t_{0})\right\Vert e^{-\frac{\lambda_{3d}}{\beta_{2}}(t-t_{0})}+\frac{\beta_{2}\iota_{4}}{\beta_{1}\lambda_{3d}}\left(1-e^{-\frac{\textbf{\ensuremath{\lambda_{3d}}}}{\beta_{2}}t}\right)}$
, for all $t\in\mathbb{R}_{\geq0}$. 
\end{thm}
\begin{IEEEproof}
Consider the candidate Lyapunov function in (\ref{eq:V}). Then using
(\ref{eq:V_Dot}), applying (\ref{eq:Analytical_Update2.2}), the
definition for $\tilde{\theta}$, and using (\ref{eq:GammaInequality}),
$\dot{\mathcal{V}}$ can be upper bounded as{\small{}
\begin{align}
\dot{\mathcal{V}} & \leq-r^{\top}k_{1}r-e^{\top}\alpha_{1}e+r^{\top}\left(R(X_{i},\tilde{\theta})+\varepsilon(X)\right)\nonumber \\
 & \;-\tilde{\theta}^{\top}\gamma_{1}\biggl(\frac{1}{2}\sum_{i=1}^{N}\Phi^{\prime\top}(X_{i},\hat{\theta})\Phi^{\prime}(X_{i},\hat{\theta})\tilde{\theta}\nonumber \\
 & \;-\sum_{i=1}^{N}\Phi^{\prime\top}(X_{i},\hat{\theta})\left(\tilde{\Delta}_{i}+R(X_{i},\hat{\theta})-R(X_{i},\tilde{\theta})+\varepsilon(X_{i})\right)\biggr)\nonumber \\
 & \;+\tilde{\theta}^{\top}\gamma_{2}\theta^{*}-\tilde{\theta}^{\top}\gamma_{2}\tilde{\theta}.\label{eq:V_Dot2Algorithm2}
\end{align}
} Using Lemma \ref{lem:ObserverLemma}, the bound $\left\Vert \tilde{\Delta}_{i}\right\Vert \leq\delta_{\Delta}$
holds for all $i\in\left\{ 1,\ldots,N\right\} $, provided $\zeta\in\mathcal{D}_{4}$,
the data is collected after the settling time $t_{\Delta}$, and the
feasibility gain condition $k_{\Delta}\Lambda_{1}>\frac{\delta_{f}^{2}}{\delta_{\Delta}^{2}}$
is satisfied. Therefore, using the bounds in (\ref{eq:Jacobian_Lagrange_Bounds_Control})
and (\ref{eq:HS_rho_bounds}), $\dot{\mathcal{V}}$ can be upper bounded
as
\begin{equation}
\dot{\mathcal{V}}\leq-\left(\lambda_{3}-\rho_{\delta}\left(\left\Vert \delta\right\Vert \right)\right)+\iota_{4},\label{eq:V_Dot3Algorithm2}
\end{equation}
when $\zeta\in\mathcal{D}_{4}$. Therefore, when $\zeta$ is initialized
such that $\zeta(t_{0})\in\mathcal{H}$, then it follows that $\textbf{\ensuremath{\lambda_{3}}}>\lambda_{3d}+\rho_{\delta}\left(\sqrt{\frac{\beta_{2}}{\beta_{1}}\left\Vert \zeta(t_{0})\right\Vert ^{2}+\frac{\beta_{2}\iota_{4}}{\beta_{1}\lambda_{3d}}}\right)$.
Recall that, because the solution $t\mapsto\zeta(t)$ is continuous,
$\zeta$ cannot instantaneously escape $\mathcal{H}$ at $t_{0}$,
therefore there exists a time-interval $\mathcal{I}_{2}$ such that
$\zeta(t)\in\mathcal{H}$ for all $t\in\mathcal{I}_{2}$, implying
\begin{equation}
\left\Vert \zeta(t)\right\Vert <\sqrt{\frac{\beta_{2}}{\beta_{1}}\left\Vert \zeta(t_{0})\right\Vert ^{2}+\frac{\beta_{2}\iota_{4}}{\beta_{1}\lambda_{3d}}}\label{eq:z_bound-1}
\end{equation}
 for all $t\in\mathcal{I}_{2}$. Because $\rho_{\delta}$ is strictly
increasing, $\rho_{\delta}\left(\left\Vert \zeta(t)\right\Vert \right)<\rho_{\delta}\left(\sqrt{\frac{\beta_{2}}{\beta_{1}}\left\Vert \zeta(t_{0})\right\Vert ^{2}+\frac{\beta_{2}\iota_{4}}{\beta_{1}\lambda_{3d}}}\right)$
for all $t\in\mathcal{I}_{2}$. As a result,
\begin{equation}
\dot{\mathcal{V}}\leq-\lambda_{3d}\left\Vert \zeta\right\Vert ^{2}+\iota_{4}\label{eq:FinalV_Algorithm2}
\end{equation}
for all $t\in\mathcal{I}_{2}.$ Using (\ref{eq:LyapBounds}), $\dot{\mathcal{V}}$
and solving the differential inequality over $\mathcal{I}_{2}$ yields
$\mathcal{V}(\zeta(t))\leq\mathcal{V}(\zeta(t_{0}))e^{-\frac{\textbf{\ensuremath{\lambda_{3d}}}}{\beta_{2}}(t-t_{0})}+\frac{\beta_{2}\iota_{4}}{\textbf{\ensuremath{\lambda_{3d}}}}\left(1-e^{-\frac{\lambda_{3d}}{\beta_{2}}t}\right)$.
Using the same process as Theorem \ref{thm:Algorithm1} it can be
shown that $\mathcal{I}_{2}$ can be extended to the interval $[t_{0},\infty)$
and $\left\Vert \zeta(t)\right\Vert <\sqrt{\frac{\beta_{2}}{\beta_{1}}\left\Vert \zeta(t_{0})\right\Vert ^{2}+\frac{\beta_{2}\iota_{4}}{\beta_{1}\lambda_{3d}}}$
for all $t\in[t_{0},\infty)$. Then, \cite[Def. 4.6]{Khalil2002}
can be invoked to conclude that $\zeta$ is bounded such that $\Vert\zeta(t)\Vert\leq\sqrt{\frac{\beta_{2}}{\beta_{1}}\Vert\zeta(t_{0})\Vert^{2}e^{-\frac{\textbf{\ensuremath{\lambda_{3d}}}}{\beta_{2}}(t-t_{0})}+\frac{\beta_{2}\iota_{4}}{\beta_{1}\lambda_{3d}}\left(1-e^{-\frac{\textbf{\ensuremath{\lambda_{3d}}}}{\beta_{2}}t}\right)}$.
Therefore, if $\zeta(t_{0})\in\mathcal{H}$, then $\zeta(t)\in\mathcal{H}\subset\mathcal{D}_{4}$
and therefore $X\in\Omega_{4}$ for all $t\geq0$. Using (\ref{eq:V})
and (\ref{eq:FinalV_Algorithm2}) implies $e,r,\tilde{\theta}\in\mathcal{L}_{\infty}.$
The fact that $x_{d},\dot{x}_{d},\ddot{x}_{d},e,r\in\mathcal{L}_{\infty}$
implies $x,\dot{x}\in\mathcal{L}_{\infty}$, using this and the fact
$\hat{\theta}\in\mathcal{L}_{\infty}$ is bounded by the use of the
projection operator implies $u$ is bounded.
\end{IEEEproof}

\section{Simulations}

Simulation results are provided to demonstrate the performance of
the developed method and comparisons are provided using the implementable
forms of the updates proposed in (\ref{eq:analytical_update1}) and
(\ref{eq:Analytical_Update2.2}) as outlined in (\ref{eq:Implementable_Update})
and (\ref{eq:Implementable_Update2}) as well as the baseline method
outlined in \cite{Patil.Le.ea2022}. To demonstrate the performance
improvement and applicability of the developed method across a variety
of nonlinear systems, simulations were performed on two unknown nonlinear
systems modeled by (\ref{eq:dynamics}) with different models for
$f(x,\dot{x})$ given by{\footnotesize{} $f_{1}(x,\dot{x})=\begin{bmatrix}\text{sin}\left(x_{1}+x_{2}\right)\text{cos}\left(\dot{x}_{1}-\dot{x}_{2}\right)+\text{cos}\left(x_{1}\right)\text{sin}\left(x_{2}\right)\text{cos}(\dot{x}_{1})\text{sin}\left(\dot{x}_{2}\right)\\
\text{cos}(x_{1})\text{sin}(x_{2})\text{cos}(\dot{x}_{1})\text{sin}\text{\ensuremath{\left(\ensuremath{\dot{x}_{2}}\right)}}-\text{sin}\left(x_{1}+x_{2}\right)\text{cos}\left(\dot{x}_{1}-\dot{x}_{2}\right)\text{sin}(x_{1})
\end{bmatrix}$ }{[}N{]} and $f_{2}(x,\dot{x})=\begin{bmatrix}x_{1}\dot{x}_{2}\text{tanh}\left(x_{2}\right)+\text{sech}^{2}\left(x_{1}\right)\\
\text{sech}^{2}\left(\dot{x}_{1}+\dot{x}_{2}\right)-\text{sech}^{2}\left(x_{2}\right)
\end{bmatrix}${[}N{]}, respectively. Simulations were run for 100s with an update
preformed every 0.01s to demonstrate the impact of the real-time adaptation.
\begin{figure}
\includegraphics[width=1.05\columnwidth]{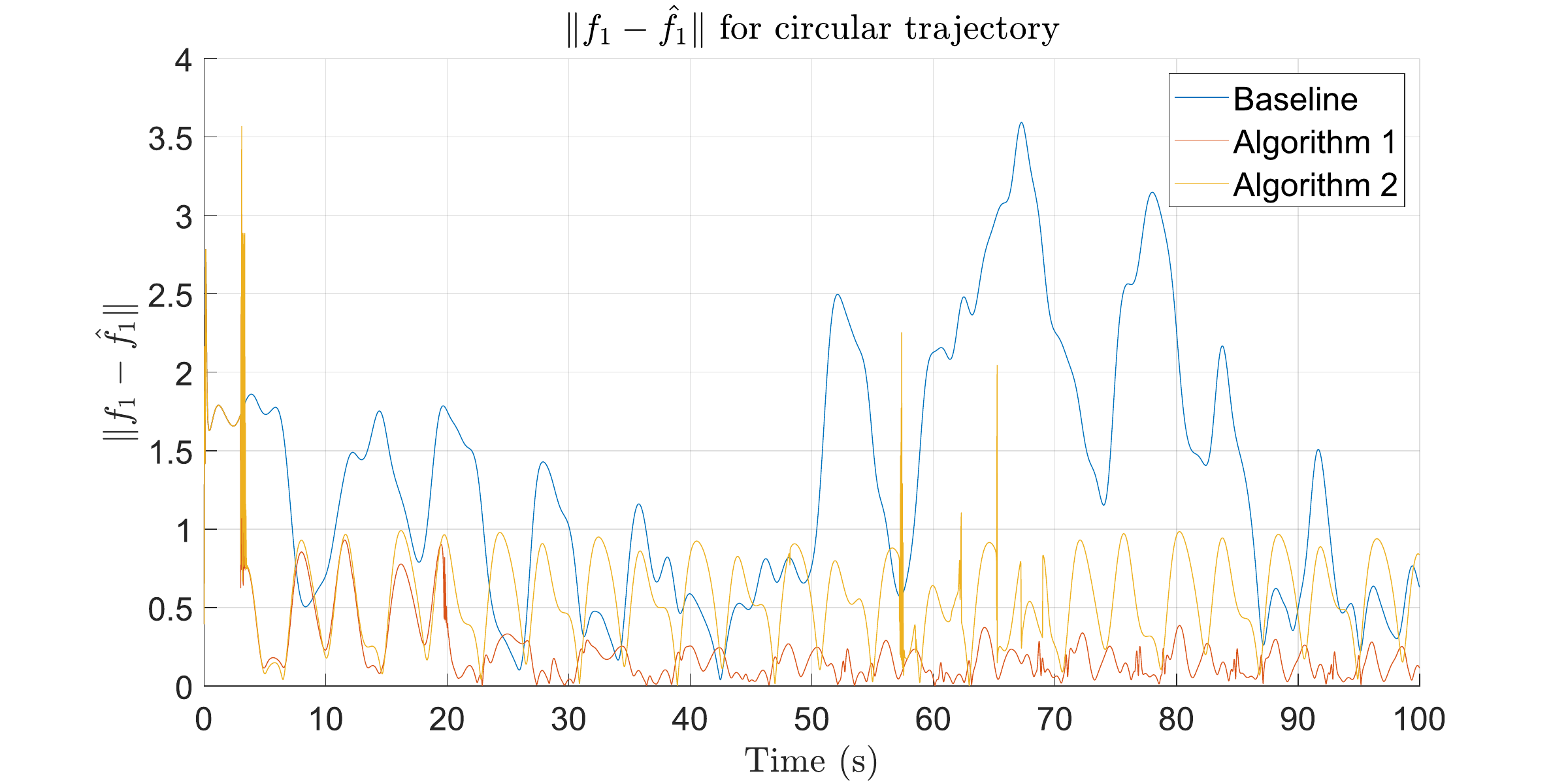}

\includegraphics[width=1.05\columnwidth]{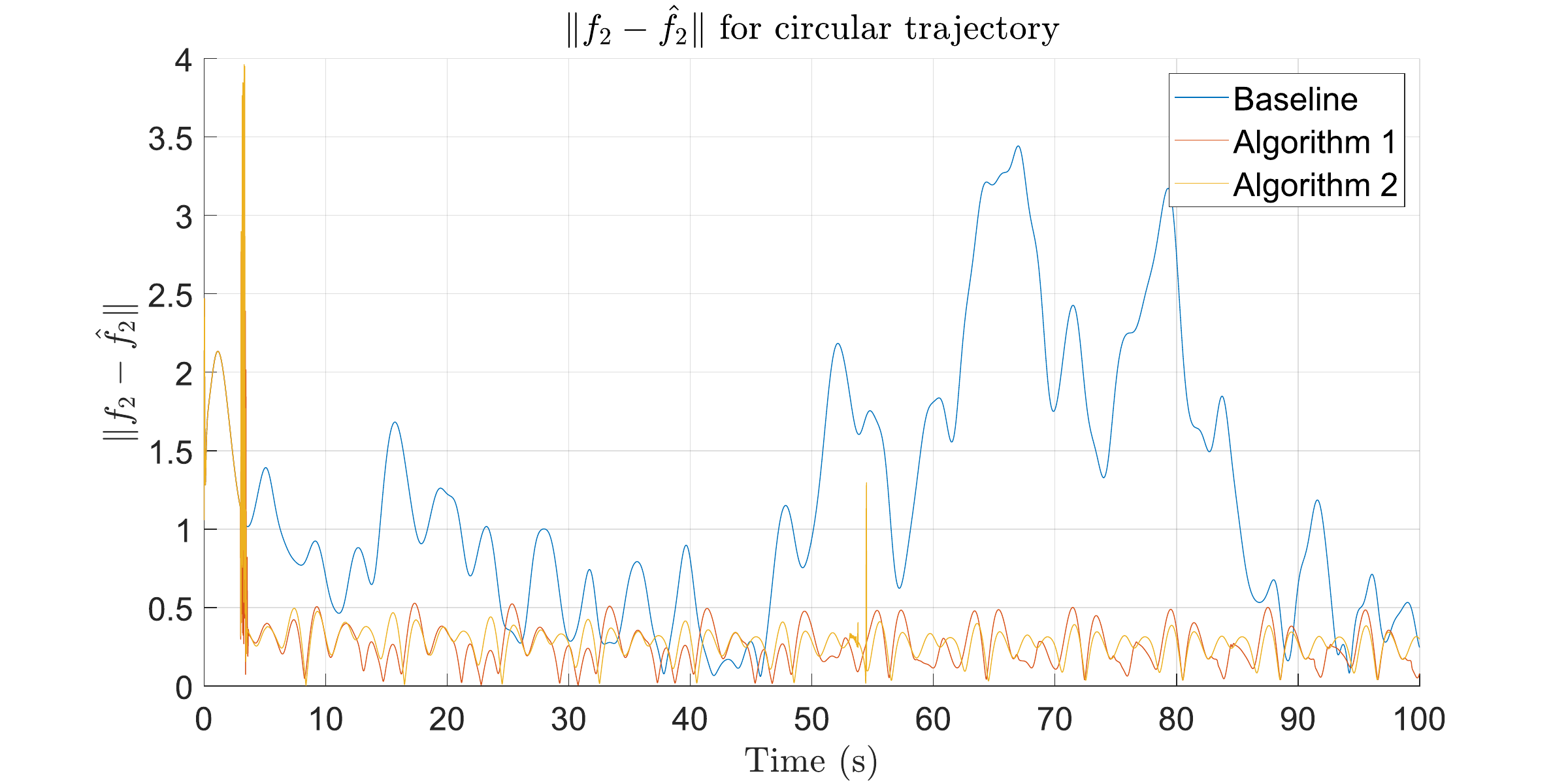}\caption{\label{fig:f_tilde_figure_circular} Function approximation error
$\Vert f(x,\dot{x})-\hat{f}(x,\dot{x})\Vert$ for circular trajectory
and $f_{1}$ and $f_{2}$ dynamics}
\end{figure}
\begin{figure}
\includegraphics[width=1.05\columnwidth]{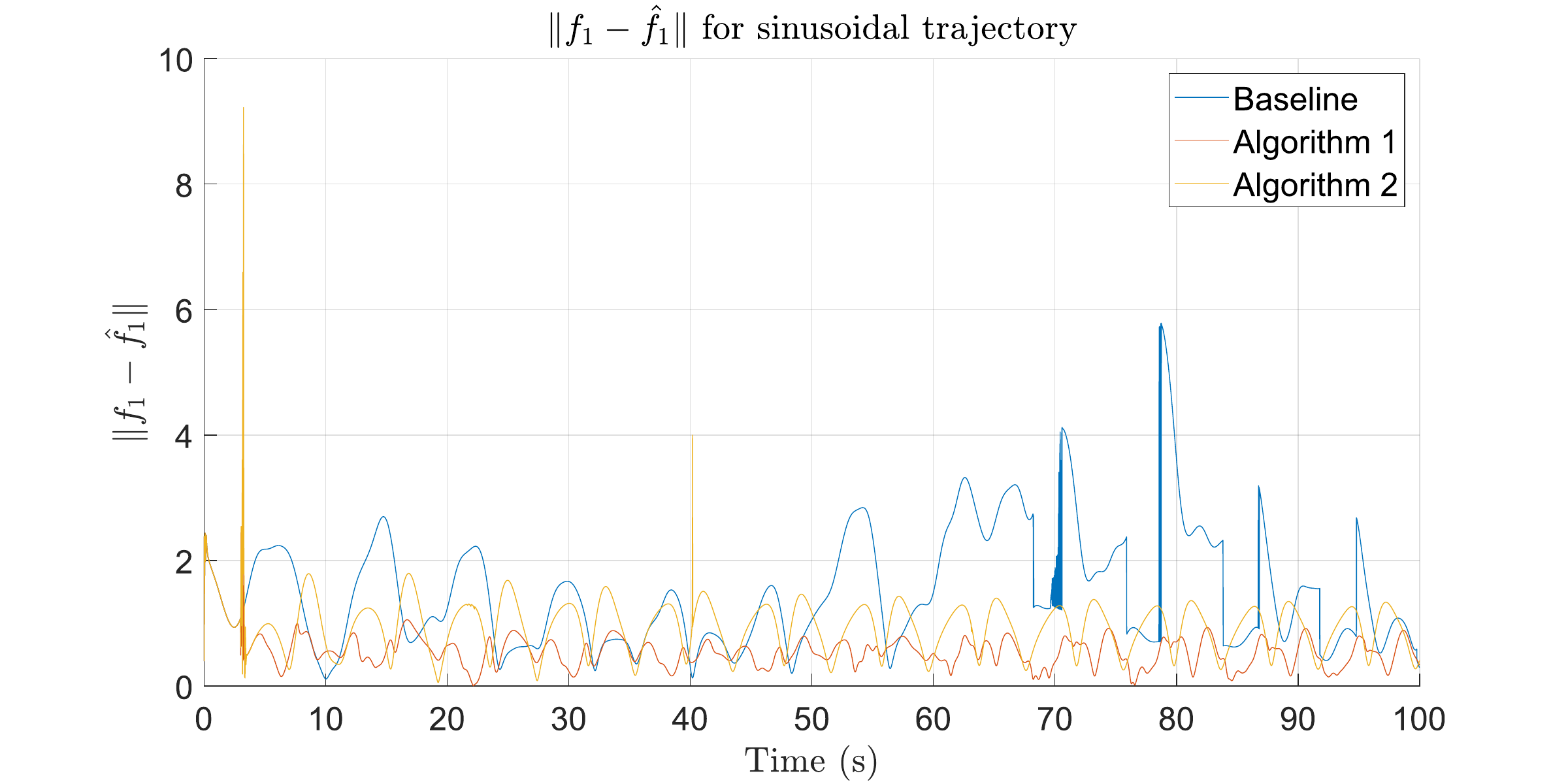}

\includegraphics[width=1.05\columnwidth]{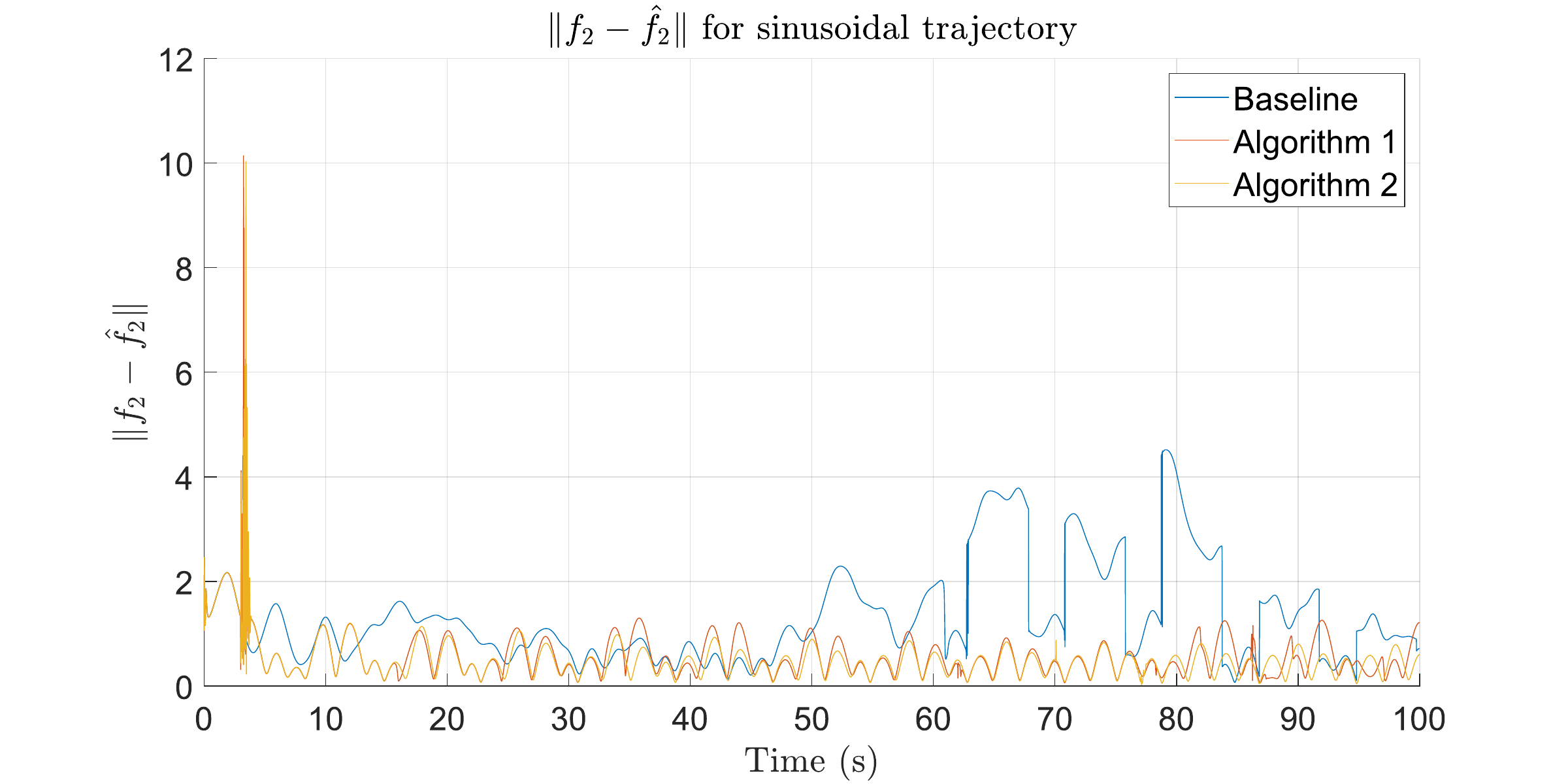}\caption{\label{fig:f_tilde_figure_sinusoid}$\Vert f(x,\dot{x})-\hat{f}(x,\dot{x})\Vert$
for Sinusoidal trajectory and $f_{1}$ and $f_{2}$ dynamics}
\end{figure}
The settling time $t_{\Delta}$ was selected as 3s and the history
stack was constructed using a sliding window of the previous 200 data
points (2s of data) with the stack being updated every 5 new data
points gathered or 0.05s. The tracking objective was preformed using
two trajectory formulations, the first was a circular desired trajectory
selected as $q_{d}=\begin{bmatrix}0.7\mathrm{cos}\left(\frac{\pi}{4}t\right)\\
0.7\text{sin}\left(\frac{\pi}{4}t\right)
\end{bmatrix}\in\mathbb{R}^{2}$ {[}rad{]} and a sinusoidal desired trajectory selected as $q_{d}=\begin{bmatrix}0.7\text{sin}\left(\frac{\pi}{4}t\right)\\
\frac{\pi}{4}\text{sin}\left(\frac{\pi}{4}t\right)
\end{bmatrix}\in\mathbb{R}^{2}$ {[}rad{]}.\textbf{ }The simulation is initialized at $q(0)=\left[1.0472,-0.5236\right]^{\top}[\text{rad}]$
and $\dot{q}(0)=\left[0,0\right][\text{rad/s]}$. To promote exploration
of the weights, an artificial disturbance described as{\footnotesize{}
$d(t)=\text{cos}^{2}(0.2t)+\text{sin}^{2}(2t)\text{cos}(0.1t)+\text{sin}^{2}(-1.2t)\text{cos}(0.5t)+\text{sin}^{2}(t)$}
was added to the update law as described in Remark \ref{rem:time varying disurbance}.
The DNNs are composed of 4 layers,\textbf{ }2\textbf{ }neurons, and
tanh activation functions. The weights of the DNN were randomly initialized
from a uniform distribution $U\left(-1,1\right)$. The gains were
selected as $\alpha_{1}=15$, $\alpha_{2}=50,$ $k_{1}=40$, $k_{\Delta}=20$,
$\beta=0.01$, $\gamma_{1}=0.12$, $\gamma_{2}=0.005$, $\gamma_{3}=0.001$,
$\Gamma(0)=1$. The same randomly selected initial weights and control
gains were used for all architectures and trajectory types to demonstrate
the performance of the adaptation under the same conditions. For the
baseline method the gain associated with the history stack in the
update law $\gamma_{1}$, and the sigma-modification technique $\gamma_{2},$
were set to be 0. Both algorithms satisfy the FE condition stated
in Assumption \ref{assum:Assumption-2.FE} a minimum of once during
the simulation run time for the circular trajectory. For the sinusoidal
trajectory, Algorithm 1 meets the FE condition while Algorithm 2 does
not, however as shown in Table \ref{tab:tracking and control effort table-Sinusoidal}
there is empirically a significant performance improvement in function
approximation error both on and off the trajectory compared to the
baseline from incorporating previously seen data in the adaptation
law.

\begin{figure}
\includegraphics[width=1.05\columnwidth]{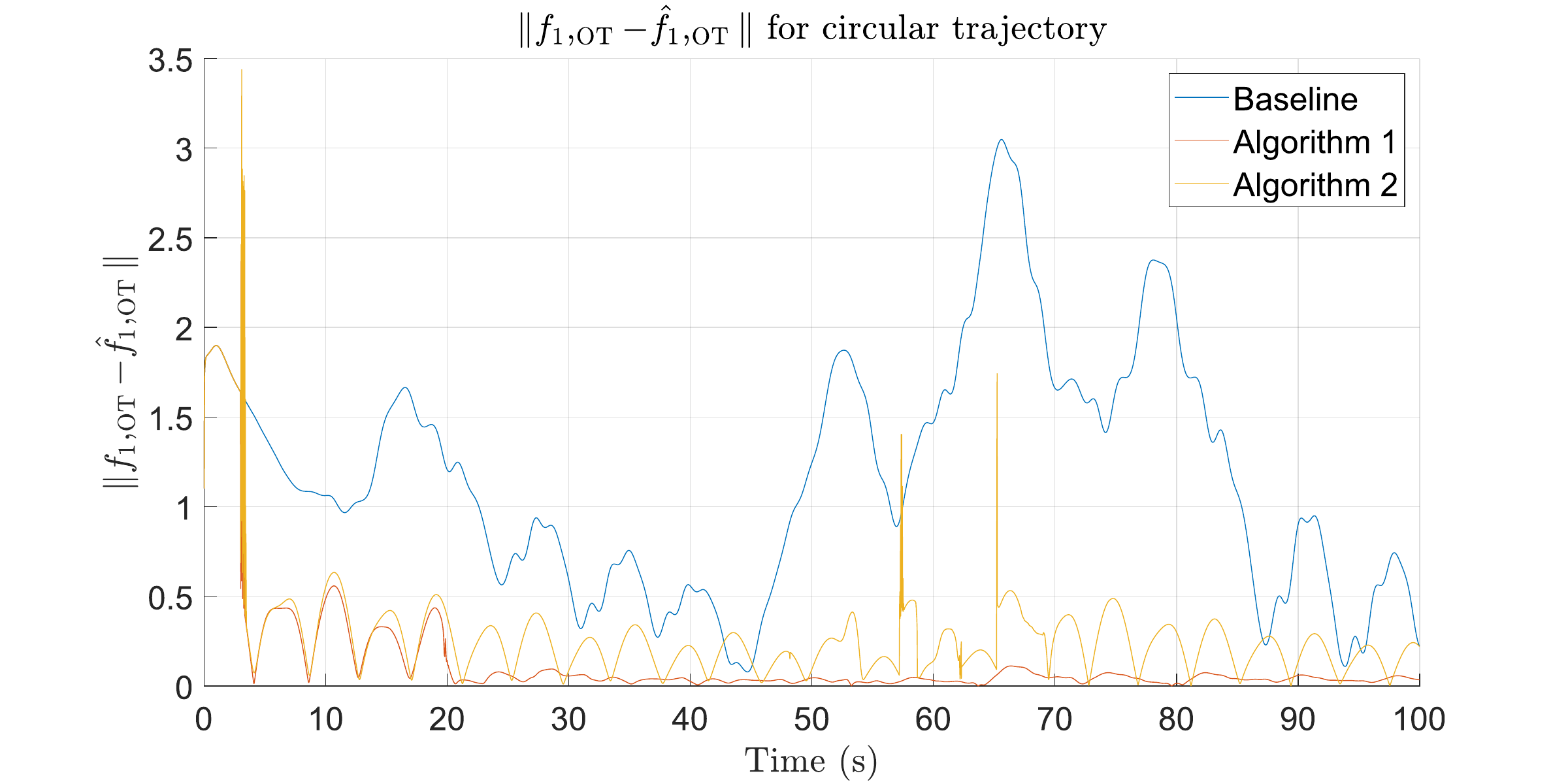}

\includegraphics[width=1.05\columnwidth]{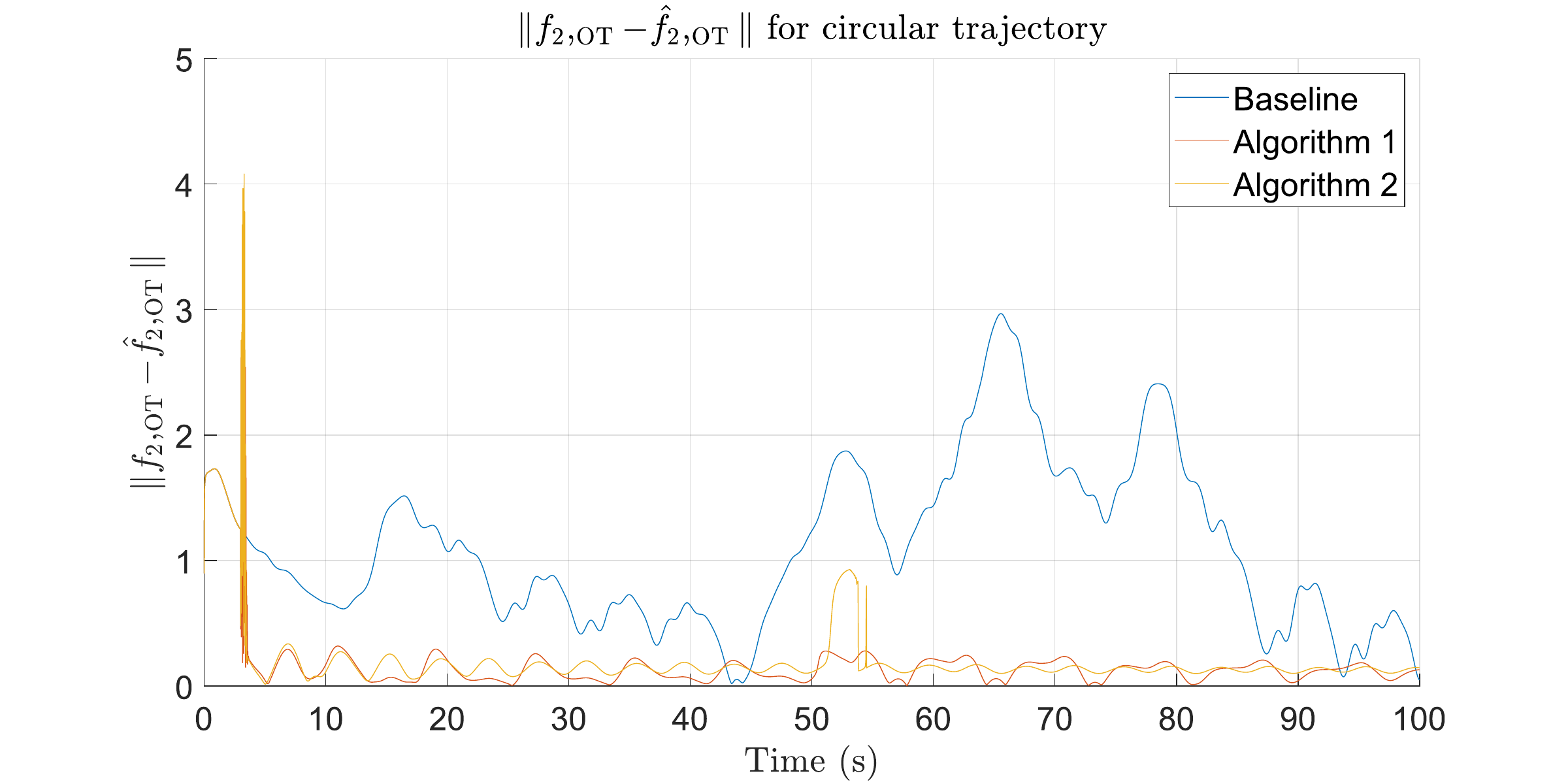}\caption{\label{fig:f_tilde_OffTraj_figureCircular}$\Vert f(x_{OT},\dot{x}_{OT})-\hat{f}(x_{OT},\dot{x}_{OT})\Vert$
for Circular trajectory and $f_{1}$ and $f_{2}$ dynamics}
\end{figure}

\begin{figure}
\includegraphics[width=1.05\columnwidth]{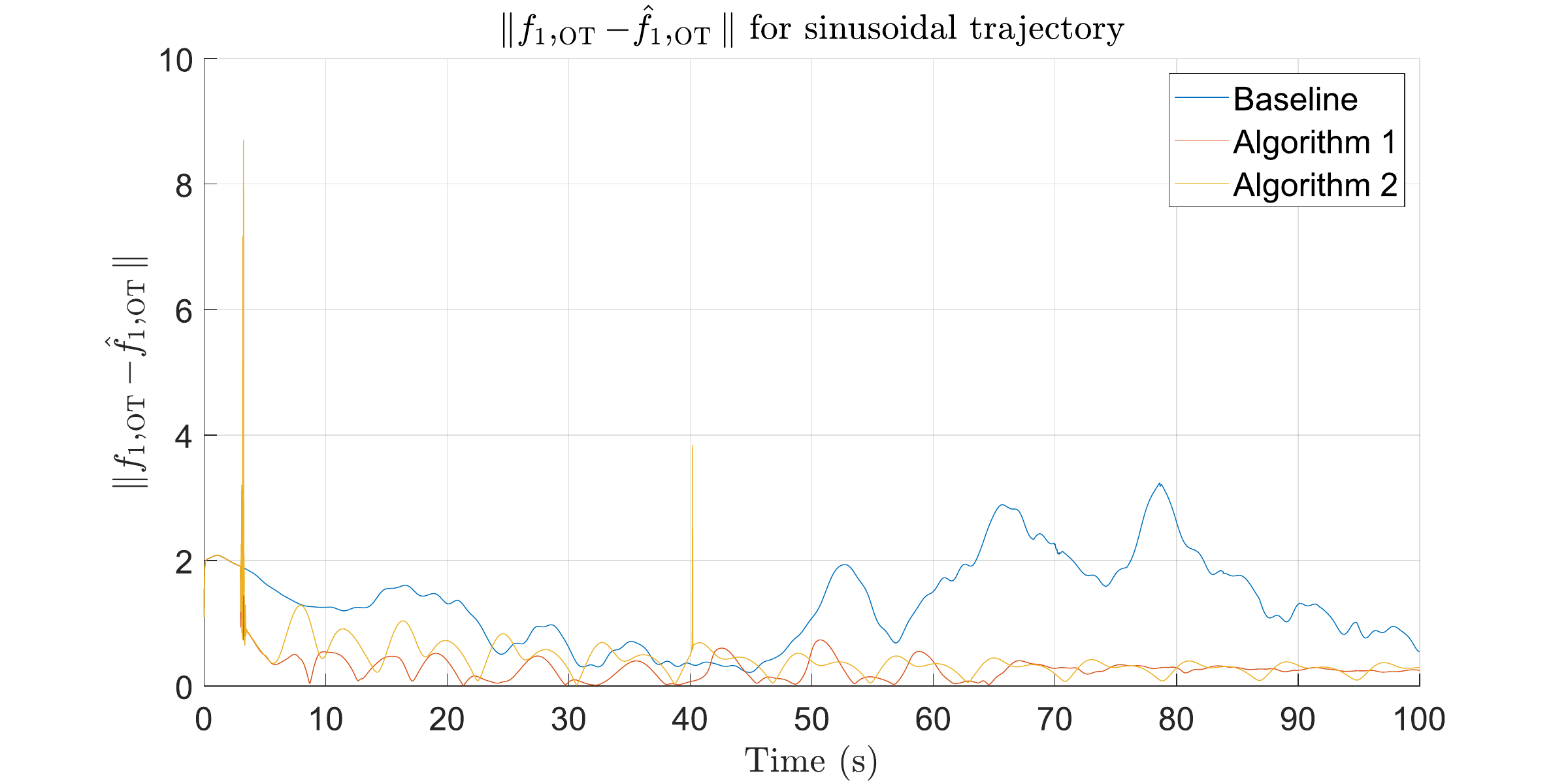}

\includegraphics[width=1.05\columnwidth]{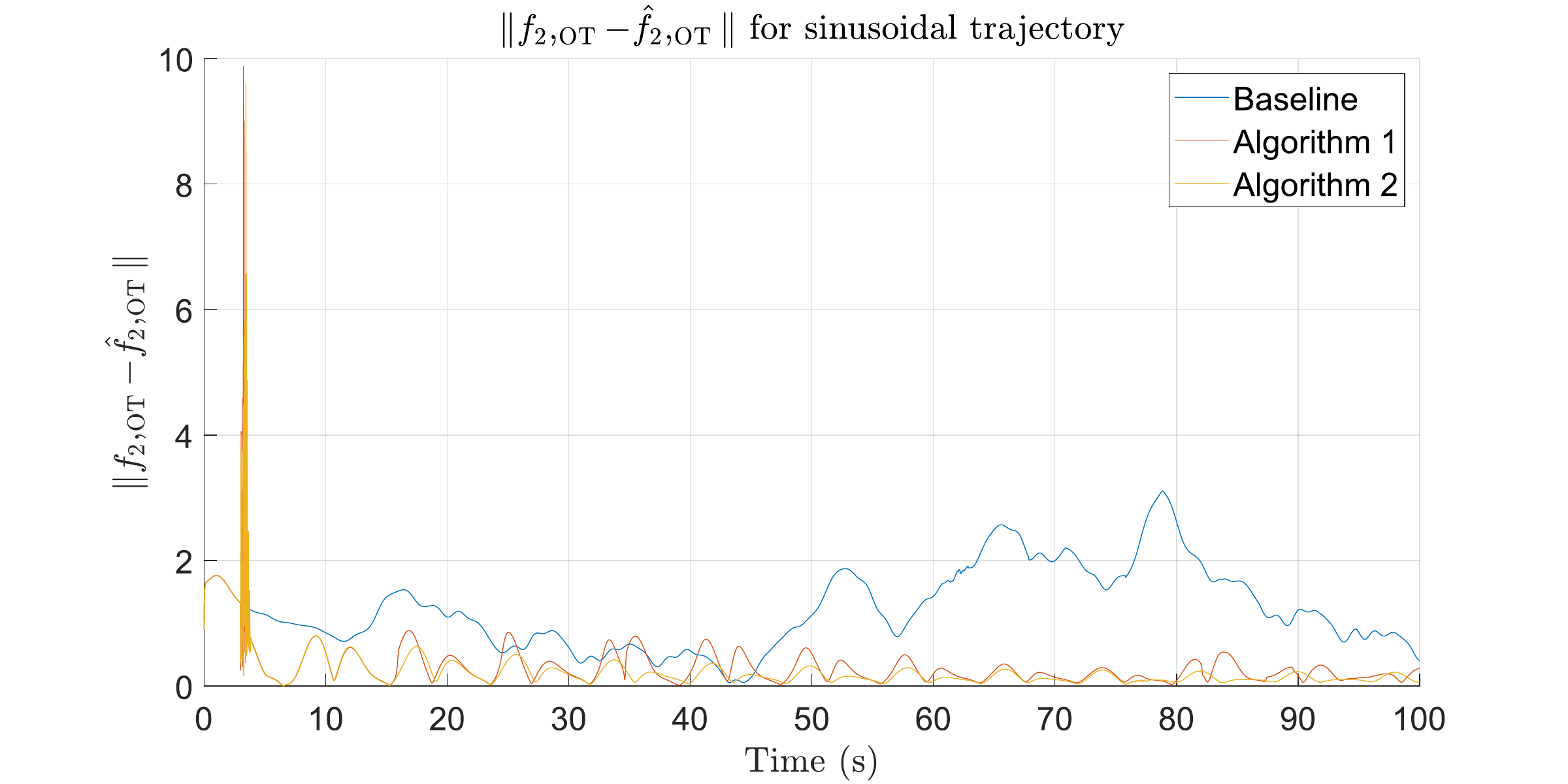}\caption{\label{fig:f_tilde_OffTraj_figureMultiSinus}$\Vert f(x_{OT},\dot{x}_{OT})-\hat{f}(x_{OT},\dot{x}_{OT})\Vert$
for Circular trajectory and $f_{1}$ and $f_{2}$ dynamics}
\end{figure}

The root mean squared (RMS) tracking error for Algorithm 1, Algorithm
2, and the baseline method are tabulated in Tables \ref{tab:tracking and control effort table-Circular}
and \ref{tab:tracking and control effort table-Sinusoidal}
\begin{table*}
\begin{centering}
\caption{\label{tab:tracking and control effort table-Circular}Tracking and
Control Effort Metrics for the Developed Methods vs. Baseline Method
for Circular Trajectory}
\centering
\par\end{centering}
\centering{}{\small{}}%
\begin{tabular}{cc|c|c|c|c|c|c}
 & \multicolumn{1}{c}{} & \multicolumn{6}{c}{Circular Trajectory}\tabularnewline
 &  & \multicolumn{2}{c|}{Algorithm 1} & \multicolumn{2}{c|}{Algorithm 2} & \multicolumn{2}{c}{Baseline}\tabularnewline
\hline 
 &  & $f_{1}$ & $f_{2}$ & $f_{1}$ & $f_{2}$ & $f_{1}$ & $f_{2}$\tabularnewline
{\small{}RMS $\Vert e\Vert$} &  & 0.0144 {[}rad{]} & 0.0144 {[}rad{]} & 0.0144 {[}rad{]} & 0.0144 {[}rad{]} & 0.0146 {[}rad{]} & 0.0146{[}rad{]}\tabularnewline
\% Improvement over Baseline &  & 1.461\% & 1.273\% & 1.307\% & 1.265\% & - & -\tabularnewline
RMS $\Vert\tau\Vert$ &  & 4.640 {[}N{]} & 4.655 {[}N{]} & 4.641 {[}N{]} & 4.656 {[}N{]} & 4.639 {[}N{]} & 4.654 {[}N{]}\tabularnewline
\% Improvement over Baseline &  & -0.0099\% & -0.0215\% & -0.0343\% & -0.0428\% & - & \tabularnewline
RMS $\Vert f(x,\dot{x})-\hat{f}(x,\dot{x})\Vert$ &  & 0.4121 {[}N{]} & 0.4269{[}N{]} & 0.6860 {[}N{]} & 0.4411 {[}N{]} & 1.562 {[}N{]} & 1.453 {[}N{]}\tabularnewline
\% Improvement over Baseline &  & 73.62\% & 70.66\% & 56.09\% & 69.64\% & - & -\tabularnewline
RMS $\Vert f(x_{OT},\dot{x}_{OT})-\hat{f}(x_{OT},\dot{x}_{OT})\Vert$ &  & 0.3518{[}N{]} & 0.3232{[}N{]} & 0.4293{[}N{]} & 0.3603 {[}N{]} & 1.548 {[}N{]} & 1.439 {[}N{]}\tabularnewline
\% Improvement over Baseline &  & 73.86\% & 74.75\% & 68.10\% & 71.85\% & - & -\tabularnewline
\end{tabular}{\small\par}
\end{table*}
\begin{table*}
\begin{centering}
\caption{\label{tab:tracking and control effort table-Sinusoidal}Tracking
and Control Effort Metrics for the Developed Methods vs. Baseline
Method for Sinusoidal Trajectory}
\par\end{centering}
\begin{centering}
\centering
\par\end{centering}
\centering{}{\small{}}%
\begin{tabular}{cc|c|c|c|c|c|c}
 &  & \multicolumn{6}{c}{Sinusoidal Trajectory}\tabularnewline
 &  & \multicolumn{2}{c|}{Algorithm 1} & \multicolumn{2}{c|}{Algorithm 2} & \multicolumn{2}{c}{Baseline}\tabularnewline
\hline 
 &  & $f_{1}$ & $f_{2}$ & $f_{1}$ & $f_{2}$ & $f_{1}$ & $f_{2}$\tabularnewline
{\small{}RMS $\Vert e\Vert$} &  & 0.0268 {[}rad{]} & 0.0268 {[}rad{]} & 0.0268 {[}rad{]} & 0.0268 {[}rad{]} & 0.0269 {[}rad{]} & 0.0269 {[}rad{]}\tabularnewline
\% Improvement over Baseline &  & 0.5408\% & 0.4217\% & 0.3816\% & 0.4205\% & - & -\tabularnewline
RMS $\Vert\tau\Vert$ &  & 7.959 {[}N{]} & 7.927 {[}N{]} & 7.961 {[}N{]} & 7.928 {[}N{]} & 7.962 {[}N{]} & 7.926 {[}N{]}\tabularnewline
\% Improvement over Baseline &  & 0.0347\% & -0.0116\% & 0.0098\% & -0.0226\% & - & -\tabularnewline
RMS $\Vert f(x,\dot{x})-\hat{f}(x,\dot{x})\Vert$ &  & 0.6210{[}N{]} & 0.7188{[}N{]} & 1.038{[}N{]} & 0.6577{[}N{]} & 1.744{[}N{]} & 1.611{[}N{]}\tabularnewline
\% Improvement over Baseline &  & 64.38\% & 55.39\% & 40.46\% & 59.19\% & - & -\tabularnewline
RMS $\Vert f(x_{OT},\dot{x}_{OT})\Vert$ &  & 0.4315{[}N{]} & 0.4878{[}N{]} & 0.5515{[}N{]} & 0.4496{[}N{]} & 1.759{[}N{]} & 1.624{[}N{]}\tabularnewline
\% Improvement over Baseline &  & 67.83\% & 64.69\% & 58.88\% & 67.46\% & - & -\tabularnewline
\end{tabular}{\small\par}
\end{table*}
 and demonstrated that all methods yielded similar tracking error
performance and required control effort. The similarity in tracking
error performance is expected due to the powerful function approximation
capabilities of DNNs. However, the developed adaptation laws are motivated
by the parameter identification objective, therefore it is expected
to see an improvement in the error between the actual function and
the estimated function which can be quantified as $f(x,\dot{x})-\hat{f}(x,\dot{x})$.
This error is quantified in Tables \ref{tab:tracking and control effort table-Circular}
and \ref{tab:tracking and control effort table-Sinusoidal} and shown
in Figures \ref{fig:f_tilde_figure_circular} and \ref{fig:f_tilde_figure_sinusoid}
indicates that the use of the developed update laws result in function
approximation performance beyond being able to compensate for the
unknown dynamics and demonstrates between a 40.5\% and 73.6\% improvement
in identifying the true function for multiple types of dynamics and
trajectories compared to a method which does not use the CL-based
update law. To evaluate the performance of the DNN from off-trajectory
data, a test data set involving 100 data points with values selected
from the distribution $U\left(-1,1\right)$ is constructed. The function
evaluated at these points is denoted $f(x_{OT},\dot{x}_{OT})$ and
the estimate at these points is denoted $\hat{f}(x_{OT},\dot{x}_{OT})$.
The mean function approximation error for the off-trajectory points
is shown in Figures \ref{fig:f_tilde_OffTraj_figureCircular} and
\ref{fig:f_tilde_OffTraj_figureMultiSinus} demonstrating improved
function approximation capabilities on unseen data for Algorithm 1
and Algorithm 2 compared to the baseline. Similarly, these capabilities
are used as a metric in Tables \ref{tab:tracking and control effort table-Circular}
and \ref{tab:tracking and control effort table-Sinusoidal}
 to demonstrate the performance improvement resulting in between 58.88\%
and 74.75\% improvement in off-trajectory function approximation compared
to the baseline method. 

\section{Conclusions}

An adaptive CL-DNN-based controller was developed for a control-affine
general nonlinear system. Leveraging previously gathered data to form
a history-stack, the developed Lb-CL adaptation is capable of achieving
both tracking error and parameter estimation error convergence. A
Lyapunov-based stability analysis guarantees ultimately bounded error
convergence for both the tracking and the weight estimation errors.
Simulations were conducted on two distinct nonlinear systems using
two different trajectories, all under the same initial conditions
and control gains. The system identification objective showed improvements
ranging from 40.5\% to 73.6\% compared to the baseline method while
providing similar tracking error performance and control effort. Similarly,
off-trajectory simulations demonstrated consistent performance gains,
with improvements of 58.88\% and 74.75\% for the developed methods,
highlighting their robust effectiveness across different systems.

\bibliographystyle{ieeetr}
\bibliography{encr,ncr,master}
\pagebreak{}
\begin{IEEEbiography}[{\includegraphics[width=1\columnwidth]{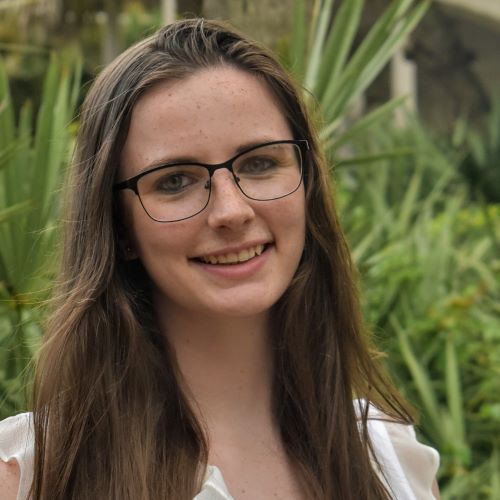}}]{Rebecca G. Hart}
 is a Ph.D. candidate in Mechanical Engineering at the University
of Florida, advised by Dr. Warren Dixon. She earned her Bachelor of
Science in Mechanical Engineering from North Carolina State University
in 2022 and her M.S. from the University of Florida in 2023. She is
a recipient of the NSF Graduate Research Fellowship. Her research
interests include Lyapunov-based control techniques, deep learning
methods, and physics-informed learning.
\end{IEEEbiography}

\vspace{-1cm}
\begin{IEEEbiography}[{\includegraphics[width=1\columnwidth]{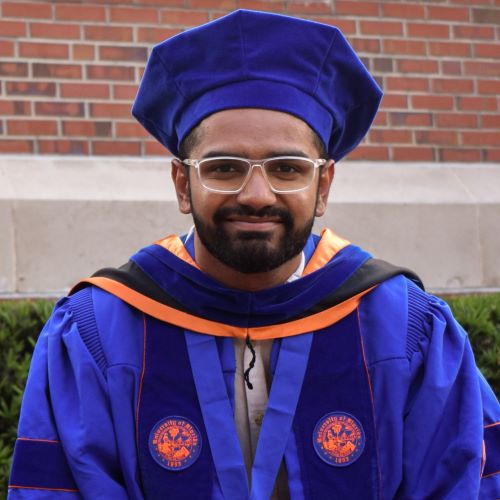}}]{Omkar Sudhir Patil}
 received his Bachelor of Technology (B.Tech.) degree in production
and industrial engineering from Indian Institute of Technology (IIT)
Delhi in 2018, where he was honored with the BOSS award for his outstanding
bachelor's thesis project. In 2019, he joined the Nonlinear Controls
and Robotics (NCR) Laboratory at the University of Florida under the
guidance of Dr. Warren Dixon to pursue his doctoral studies. Omkar
received his Master of Science (M.S.) degree in mechanical engineering
in 2022 and Ph.D. in mechanical engineering in 2023 from the University
of Florida. During his Ph.D. studies, he was awarded the Graduate
Student Research Award for outstanding research. In 2023, he started
working as a postdoctoral research associate at NCR Laboratory, University
of Florida. His research focuses on the development and application
of innovative Lyapunov-based nonlinear, robust, and adaptive control
techniques.
\end{IEEEbiography}

\vspace{-1cm}
\begin{IEEEbiography}[{\includegraphics[width=1\columnwidth]{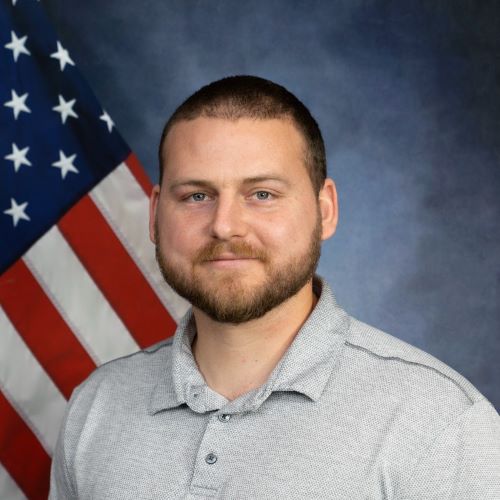}}]{Zachary I. Bell}
 received his Ph.D. from the University of Florida in 2019 and is
a researcher for the Air Force Research Lab. His research focuses
on cooperative guidance and control, computer vision, adaptive control,
and reinforcement learning.
\end{IEEEbiography}

\vspace{-1cm}
\begin{IEEEbiography}[{\includegraphics[width=1\columnwidth]{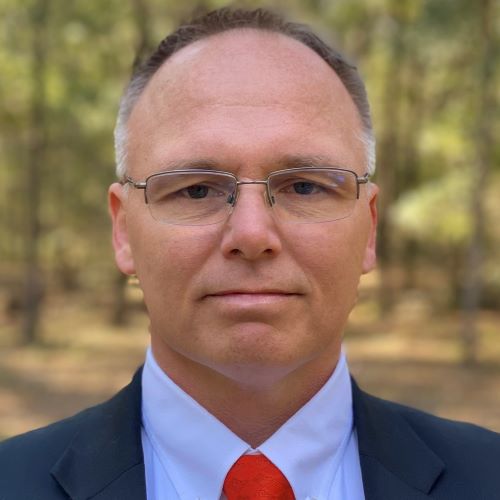}}]{Prof. Warren Dixon}
 received his Ph.D. in 2000 from the Department of Electrical and
Computer Engineering from Clemson University. He worked as a research
staff member and Eugene P. Wigner Fellow at Oak Ridge National Laboratory
(ORNL) until 2004, when he joined the University of Florida in the
Mechanical and Aerospace Engineering Department. His main research
interest has been the development and application of Lyapunov-based
control techniques for uncertain nonlinear systems. His work has been
recognized by the 2019 IEEE Control Systems Technology Award, (2017-2018
\& 2012-2013) University of Florida College of Engineering Doctoral
Dissertation Mentoring Award, 2015 \& 2009 American Automatic Control
Council (AACC) O. Hugo Schuck (Best Paper) Award, the 2013 Fred Ellersick
Award for Best Overall MILCOM Paper, the 2011 American Society of
Mechanical Engineers (ASME) Dynamics Systems and Control Division
Outstanding Young Investigator Award, the 2006 IEEE Robotics and Automation
Society (RAS) Early Academic Career Award, an NSF CAREER Award (2006-2011),
the 2004 Department of Energy Outstanding Mentor Award, and the 2001
ORNL Early Career Award for Engineering Achievement. He is an ASME
Fellow (2016) and IEEE Fellow (2016), was an IEEE Control Systems
Society (CSS) Distinguished Lecturer (2013-2018), served as the Director
of Operations for the Executive Committee of the IEEE CSS Board of
Governors (BOG) (2012-2015), and served as an elected member of the
IEEE CSS BOG (2019-2020). His technical contributions and service
to the IEEE CSS were recognized by the IEEE CSS Distinguished Member
Award (2020). He was awarded the Air Force Commander's Public Service
Award (2016) for his contributions to the U.S. Air Force Science Advisory
Board. He is currently or formerly an associate editor for ASME Journal
of Journal of Dynamic Systems, Measurement and Control, Automatica,
IEEE Control Systems, IEEE Transactions on Systems Man and Cybernetics:
Part B Cybernetics, and the International Journal of Robust and Nonlinear
Control.

\enlargethispage{-12.5cm}
\end{IEEEbiography}

\end{document}